\documentclass[a4paper,fleqn,usenatbib]{mnras}
\usepackage{newtxtext,newtxmath}

\usepackage[T1]{fontenc}
\usepackage{ae,aecompl}
\usepackage{graphicx}	
\usepackage{amsmath}	
\usepackage{amssymb}	
\usepackage{epstopdf}

\usepackage{ulem}
\usepackage{color} 
\usepackage{url}

\newcommand{\angstrom}{\textup{\AA}}

\title[Comparison of different population models and stellar libraries]{Recovering stellar population parameters via different population models and stellar libraries}
\author[J. Q. Ge]{
Junqiang Ge, $^1$\thanks{E-mail: jqge@nao.cas.cn}
Shude Mao,$^{2,1}$
Youjun Lu, $^{1,3}$
Michele Cappellari, $^4$
and Renbin Yan $^5$\\
$^{1}$National Astronomical Observatories, Chinese Academy
  of Sciences, 20 Datun Road, Beijing, 100101, China \\
$^{2}$Physics Department and Tsinghua Center for Astrophysics,
               Tsinghua University, Beijing, 100084, China\\
$^{3}$School of Astronomy and Space sciences, University of Chinese Academy
          of Sciences, 19A Yuquan Road, Beijing, 100049, China\\
$^{4}$Sub-Department of Astrophysics, Department of Physics,
          University of Oxford, Denys Wilkinson Building, Keble Road,
          Oxford, OX1 3RH, UK\\	
$^{5}$Department of Physics and Astronomy, University of Kentucky,
             505 Rose Street, Lexington, KY 40506, USA\\
}

\date{Accepted XXX. Received YYY; in original form ZZZ}

\pubyear{2018}

\begin{document}
\label{firstpage}
\pagerange{\pageref{firstpage}--\pageref{lastpage}}
\maketitle

\begin{abstract}
Three basic ingredients are required to generate a simple stellar population (SSP)
library, i.e., an initial mass function (IMF), a stellar evolution model/isochrones, 
and an empirical/theoretical stellar spectral library. However, there are still some
uncertainties to the determination and understanding of these ingredients. 
We perform the spectral fitting to test the relative parameter offsets between 
these uncertainties using two different stellar population 
models, two different empirical stellar libraries, two different isochrones, and the 
Salpeter and Chabrier IMFs. Based on these setups, we select five SSP libraries generated
with the Galaxev/STELIB and Vazdekis/MILES models, and apply them to the pPXF full-spectrum fitting of 
both MaNGA and mock spectra. 
We find that: 1) Compared to the Galaxev/STELIB model, spectral fitting qualities with the 
Vazdekis/MILES model have significant improvements for those metal-rich (especially over-solar) 
spectra, which cause better reduced $\chi^2$ distributions and more precisely fitted absorption lines. 
This might due to the lack 
of metal rich stars in the empirical STELIB library, or code improvement of the Vazdekis model.
2) When applying the Vazdekis/MILES model for spectral fitting, the IMF variation will lead to 
not only a systematic offset in $M_*/L_r$, but also offsets in age and metallicity, and these offsets 
increase with increasing stellar population ages.
However, the IMF-variation caused metallicity offsets disappear in the case of Galaxev/STELIB based libraries.
3) The Padova2000 model provides a better match to the MaNGA galaxy spectra at [M/H]$_L<-1.0$, 
while the BaSTI model match the local galaxy spectra better at [M/H]$_L>-1.0$.
Current tests suggest that spectral fitting with the Vazdekis/MILES+BaSTI combination would be a better choice 
for local galaxies.
\end{abstract}




\section{Introduction}
\label{sec:introduction}
Galaxy formation and evolution processes can partly be resolved by analyzing their stellar 
population properties, e.g. star formation history (SFH), stellar mass, age, metallicity 
and abundance pattern, dust extinction, and gas fraction \citep[see a review by][]{conroy2013}. 
However, many of these population parameters are encoded together and need to be decomposed. 
One way of the decomposition is by stellar population synthesis modelling \citep[]{tinsley1968},
which is initially developed at a certain age and metallicity 
\citep[e.g.][]{tinsley1968, ssb1973, tg1976, bruzual1983}, and then
improved to the whole age and metallicity parameter spaces \citep[][]{cb1991, bc1993, 
bcf1994, worthey1994, frv1997, maraston1998, leitherer1999, vazdekis1999, walcher2011}.

Based on the stellar population synthesis modelling, one can produce a single stellar population (SSP) library by assuming 
a certain initial mass function (IMF), a stellar evolution model/isochrones, and an empirical stellar spectral 
library. Together with a star formation history, chemical evolution, and dust attenuation and emission 
model to the SSP library, one can then construct composite stellar population (CSP) templates.

For the three ingredients required for generating a SSP library, i.e., IMF, isochrones, and empirical stellar spectral library,
many efforts have been dedicated to their measurements or improvements. 

The IMF was initially described with the form $dN/dM \propto M^{-\alpha}$ with $\alpha=2.35$ \citep{salpeter1955}.
According to the measurements from solar neighborhood, the form deviates from the Salpeter form only at $M_*<M_{\sun}$
with a log-normal or shallower power law \citep{Kroupa2001, Chabrier2003}. Historically no strong IMF variation evidence
is found before 2010 \citep[e.g.][]{belldeJong2001, Bastian2010}. Since then, different methods are proposed 
and the derived results support a non-universal IMF at the low mass end \citep[see a review by][]{Cappellari2016}. 
Evidence is obtained mainly in three ways: 1) gravitational lenses combined with dynamical modelling 
\citep[e.g.][]{Auger2010, Treu2010, Posacki2015} or spatially resolved galaxy kinematics and dynamics 
\citep[e.g.][]{cappellari2012, Lasker2013, Tortora2013, Posacki2015, li2017};
2) analyses of IMF-sensitive absorption-line indicators based on stellar population 
modelling \citep[][]{vC2010, cv2012, Spiniello2012, LaBarbera2013, Spiniello2015, LaBarbera2016, 
Zieleniewski2017, Parikh2018, Vaughan2018}.

A number of public isochrone tables are now available \citep[e.g.][]{Schaller1992, bertelli1994, 
Cassisi2000, girardi2000, Yi2001, Cioni2006, Dotter2007}. The calculation of isochrones from different groups focus
on different aspects of stellar evolution \citep[see the summary by][]{conroy2013}. 
The Geneva models \citep[][]{schaller1992, mm2000} trace the evolution of high-mass stars well 
but has no isochrones for low-mass stars. The $Y^2$ models \citep{Yi2001, Yi2003}, 
the Dartmouth models \citep{Dotter2008}, and the Victoria-Regina models \citep{VB1985, VBD2006}
focus on the evolution of stars in the main sequence, red giant and horizontal branches. 
The Lyon models \citep{CB1997, Baraffe1998} are widely used in tracing 
isochrones of very low-mass stars and brown dwarfs.
The most popular models can span a wide 
range in ages, metallicities and evolutionary phases. These include the
Padova \citep[][]{bertelli1994,girardi2000, marigo2008} and BaSTI \citep[][]{pietrinferni2004, cordier2007} 
models. 

The third ingredient for producing a SSP library is stellar spectral library, 
which includes both theoretical ones, such as the BaSel library \citep[][]{Lejeune1997, Lejeune1998, Westera2002}, 
and empirical ones, including \cite{gs1983},
\cite{pickles1998}, \cite{jones1999}, ELODIE \citep{ps2001}, STELIB \citep{leborgne2003},
Indo-US \citep{valdes2004}, GRANADA \citep{Martins2005}, NGSL \citep{gregg2006, hl2011},
MILES \citep{sb2006}, IRTF \citep{rayner2009},
and the X-shooter library \citep{chen2011}.

With different combinations of the above three ingredients, there are many public SSP libraries available, 
such as BC03 \citep{bc2003}, M05 \citep{Maraston2005}, FSPS \citep[][]{cgw2009}, Vazdekis/MILES 
\citep{vazdekis2010}, and M11 \citep{ms2011}.

Many works have been devoted to identify the advantage and weakness of different SSP libraries.
\cite{Koleva2008} compared the BC03 and Vazdekis/MILES SSP libraries by concentrating on old population of Galactic 
globular clusters, and found that BC03 library had systematic
biases at non-solar metallicity, which might be due to the poor metallicity coverage of the STELIB empirical 
stellar spectral library.
\cite{CG2010} applied BC03, M05, and FSPS SSP libraries to multi-band photometric data, 
and summarized that these libraries had different performance in different comparison cases.
\cite{GC2010} applied different combinations of isochrones and empirical stellar libraries to 
full-spectrum fitting of star clusters, and concluded that the BC03 library may underestimate metallicity
by as much as 0.6 dex, and overestimate extinction by 0.1 mag. 
If one uses the SSP library with MILES and GRANADA empirical
stellar spectral library, the fitting accuracies could be improved by 0.1 dex in age and 0.3 dex in metallicity.
\cite{Chen2010} analyzed full-spectrum fitting results of SDSS galaxies in different types, and derived significant difference
when applying different SSP libraries using the STARLIGHT fitting.

Star clusters or globular clusters, which (mostly) have a single-age and single-metallicity, are considered 
as good targets for validation of SSP libraries.
When observing galaxies with an integral field unit (IFU) instrument, we can spatially resolve them and obtain spectra
at different radii. Those spatially resolved spectra from different types of galaxies, which actually include 
many galaxy evolutionary phases and cover a wide range of ages and metallicities, can also be used for validating 
the advantage and disadvantage of different SSP libraries.

Over the past two decades, many IFU surveys are designed based on different science goals.
There are six IFU surveys focusing on a certain type of galaxies.
The SAURON \citep{deZeeuw2002} and ATLAS$^{\rm 3D}$ \citep{Cappellari2011} performed IFU 
survey of local elliptical and lenticular galaxies. The DiskMass Survey \citep{Bershady2010}
targets 146 face-on star-forming galaxies at high spectral resolution (R $\sim$ 10,000).
The VENGA survey \citep{Blanc2013} points to 30 nearby spiral galaxies with high spatial resolution 
and deep observations. The SLUGGS Survey \citep{Brodie2014} focuses on 25 nearby early-type galaxies
with large spatial coverage (reach to $\sim$8 Re). The MASSIVE Survey \citep{Ma2014} observes
$\sim$100 most massive (stellar mass $M_* > 10^{11.5} M_{\sun}$) galaxies within 108 Mpc.
There are also three surveys targeting all types of galaxies.
The CALIFA survey \citep{Sanchez2012} has already finished their $\sim$600 galaxy survey with two sets of
wavelength coverages and spectral resolution: V500 (3745--7500\AA, R=850) and V1200 (3400--4840\AA, R=1650).
The SAMI survey \citep{Croom2012} will observe 3400 galaxies with two wavelength channels 
(3700-5700 \AA~ and 6250-7350\AA). The largest of all, the SDSS IV/MaNGA survey \citep{Bundy2015} aims to 
observe 10,000 galaxies by covering $10^9M_{\sun}<M_*<10^{11}M_{\sun}$
and redshift $z<0.15$ with spectral resolution $R\sim 2000$ and a wide wavelength coverage $3600-10300$\AA.

With these IFU data, we can make further progress on the understanding of galaxy evolution.
However, to extract stellar population parameters, one has a bewildering
number of options in the SSPs, and is pressed to understand the relative strengths and weaknesses 
in different routines.

In \cite{ge2018}, we studied the spectral fitting biases and scatters of two 
full-spectrum fitting algorithms, pPXF and STARLIGHT, in the absence of 
any model uncertainties. We found that pPXF gave stellar population parameters with smaller biases 
and scatters than STARLIGHT. We also presented the relationships between parameter biases (or scatters) 
and the signal-to-noise ratio (S/N) of the spectra.
By dividing the population parameter results into different metallicity bins and presenting population parameters at different ages, 
we explained the fitting biases and scatters more clearly than using the real data.

With the clear interpretation of algorithm-induced biases and scatters, we are going to 
analyze how the resulting population parameters will be affected by the assumptions of stellar evolution model, 
initial mass function, and stellar spectral library through both simulations and observations.
In this paper, we focus on fitting offsets\footnote{Here we use the word ``offset" to describe those 
parameter differences fitted with different SSP libraries, since we do not know which one is true. 
While we use ``bias'' to desribe the difference between the fitted parameter and its true value, e.g., simulations in Appendix \ref{simu_bias}.} derived from different SSP libraries, 
by dividing the measured luminosity-weighted metallicities into different bins, then present the parameter difference
caused by different SSP libraries based on both observation and simulation.

We will select the full-spectrum fitting code and SSP libraries, and describe our spectral fitting procedures in Section 2. 
In Section 3, we analyze the SDSS-IV/MaNGA data with our selected SSP 
libraries, and compare the relative parameter offsets induced by the two choices of each ingredient. 
To have a direct judgement on which model is closer to the reality, we do the same analyses with simulation to
interpret the origin of these relative offsets.
We discuss the scatter of our results, and also compare them with previous works in Section 4. 
The conclusion is made in Section 5, in which we also give 
recommendation on how to select these ingredients for real spectral fitting.

\section{Preparation for model uncertainty tests}
\subsection{Full-spectrum fitting code: pPXF}

\cite{ge2018} have performed a systematic analysis on the algorithm bias and scatter of both 
pPXF \citep{CappellariEmsellem2004, Cappellari2017} and STARLIGHT \citep{cid2005} full-spectrum fitting 
codes, and found that pPXF can recover stellar population parameters better than STARLIGHT and have 
almost no parameter biases on dust extinction and error spectrum types. More recently, \cite{cid2018} 
showed that the convergence of STARLIGHT can be improved with a different choice of the Markov chain 
initialization in the code. But this sensitivity to the initial conditions confirms its intrinsic 
weakness. In any case, even at comparable accuracy, \cite{ge2018} found pPXF to be two to three 
orders of magnitude faster than STARLIGHT, and this alone would justify our choice of pPXF in this 
paper. Therefore, in this work, we apply pPXF (Python version 
V6.7.6\footnote{Available from \url{http://purl.org/cappellari/software}}) for model uncertainty tests.

\subsection{Empirical stellar spectral libraries}
As is compared by \cite{vazdekis2010}, the STELIB empirical stellar spectral library shows almost the poorest
coverage of fundamental parameter space, while the MILES library is the richest one. To have a thorough analysis 
on the effect of model uncertainties caused by empirical stellar spectral library, we select STELIB and
MILES libraries for comparisons. In principle, the one with more stars should represent the real case better.

The STELIB library includes 245 stellar spectra with wavelength range 3200--9500\AA~ and spectral 
resolution $R\sim 2000$ \citep{leborgne2003}. This library has a shortage of hot ($>10,000$K), metal-rich, 
and cool dwarf stars.

The MILES library contains 985 stars with wavelength coverage 
from 3500 to 7500\AA~ and spectral resolution FWHM=2.3\AA~ \citep{sb2006}.
Compared to STELIB library, the MILES library improved the range of metallicity coverage, number of giant stars, 
and also the flux and wavelength calibrations.

\subsection{IMF selection}
\begin{figure*}
\centering
\includegraphics[angle=0.0,scale=0.65,origin=lb]{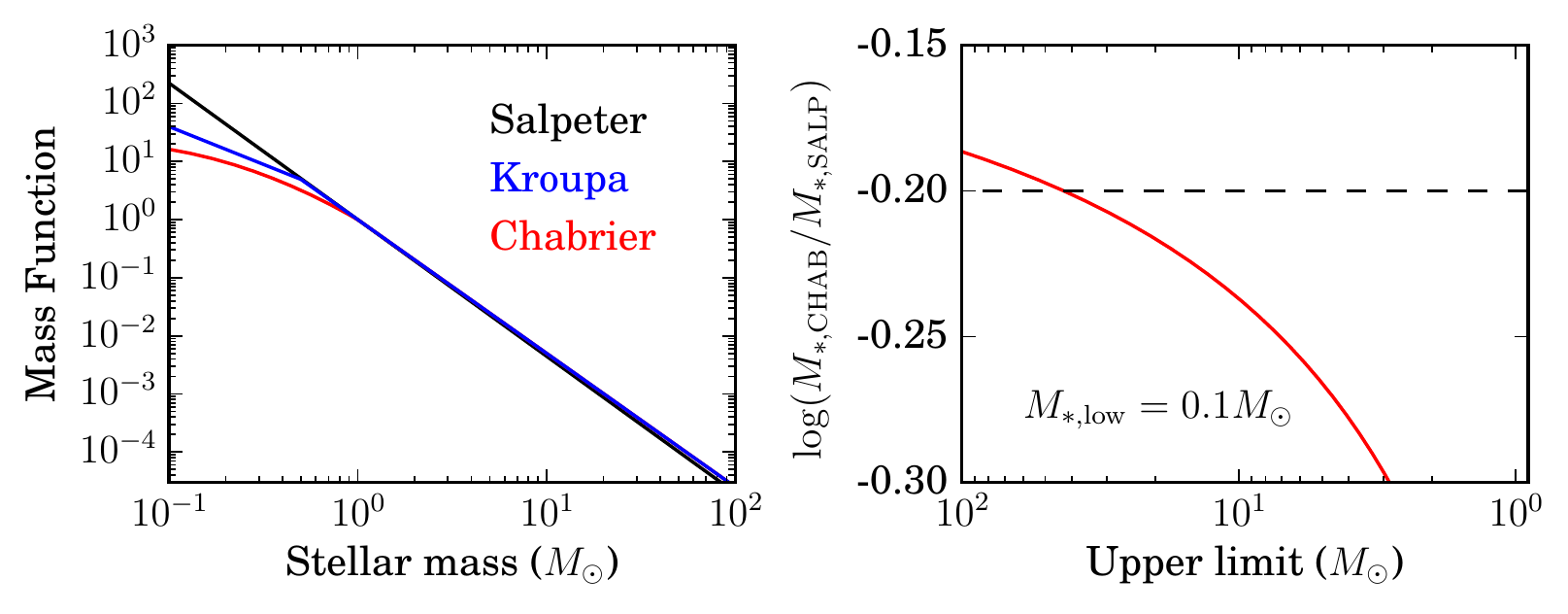}
\includegraphics[angle=0.0,scale=0.65,origin=lb]{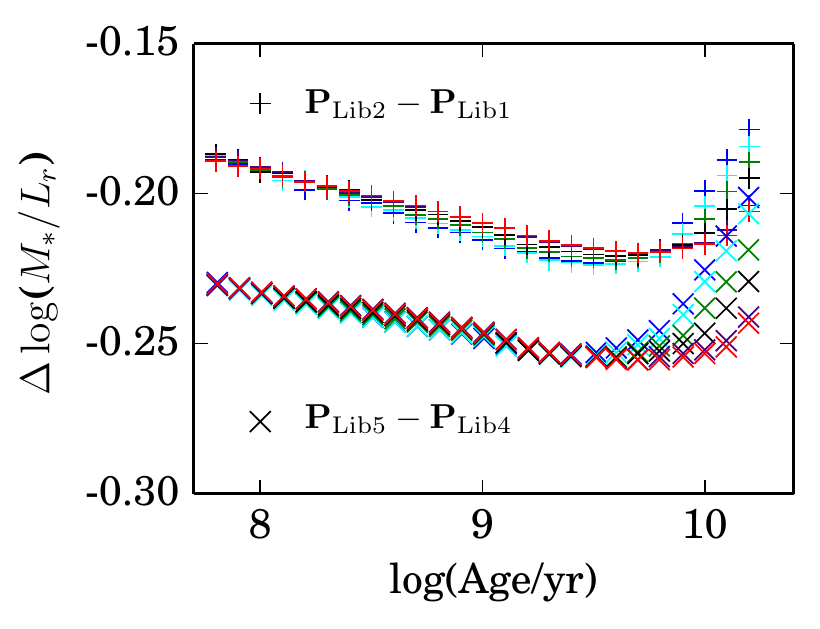}
\caption{
The Salpeter (black color), Kroupa (blue color), and Chabrier (red color) IMFs are shown in the left panel.
With the stellar evolution, high mass stars explode more and more, and the $M_*/L$ difference between
the Salpeter and Chabrier IMFs becomes larger. The middle panel shows the integrated stellar mass (integrating from
0.1$M_{\sun}$ to the upper limit) difference between the Chabrier and Salpeter IMFs. The right panel presents 
the $\Delta M_*/L_r$ between Chabrier and Salpeter IMFs along the stellar evolution time for Vazdekis/MILES 
(plus symbols) and Galaxev/STELIB (cross symbols) models, respectively.}
\label{imf}
\end{figure*}

Currently, there are mainly three kinds of IMFs calibrated based on the stellar observations of our Milky Way:
Salpeter, Kroupa and Chabrier (Figure \ref{imf} left). The Salpeter IMF is a power-law,
while the Kroupa and Chabrier IMFs reduce the contribution of low mass stars ($m_*<1 M_{\sun}$) at different levels.
The Salpeter IMF has the largest fraction of low mass stars, leading to the largest $M_*/L$ in the optical 
at a certain age and metallicity among the three IMFs.
While the Chabrier IMF yields the smallest $M_*/L$. We hence select the Salpeter and Chabrier IMFs for analyses. 
The results for the Kroupa IMF would be in between.

The integrated stellar mass difference between the Salpeter and Chabrier IMFs as a function of upper mass 
limit is shown in the middle panel of Figure \ref{imf}. With stellar evolution, high mass stars explode 
first, hence the total mass difference between the two IMFs increases with decreasing upper mass limit. 
Here we only calculate the mass of stars without including the mass loss from stellar wind, 
and also exclude the mass contribution from stellar remnants, including white dwarfs, neutron stars, 
and black holes.

At the beginning of stellar evolution, the integrated stellar mass of the Chabrier IMF is
0.15 dex lower than that of Salpeter's IMF. The integrated mass difference then increases with time, due 
to the decreased fraction of high mass stars. However, when we calculate the $M_*/L$,
the $M_*$ contains not only living stars, but also stellar remnants. Therefore, the $M_*/L$ difference caused by 
the two IMFs will not decrease as steep as shown in the middle panel of Figure \ref{imf}. 
For all the calculations in our work, the $M_*/L$ means $M_{*+remn.}/L$.
After taking the stellar remnants into account, the $\Delta M_*/L$ will not simply decrease along 
the stellar evolution time, but have a turnover after 3 Gyr (the right panel of Figure \ref{imf}). 
The trends of the $\Delta M_*/L$ vs. stellar evolution time depend on different stellar evolution 
isochrones and steller population synthesis models.

\subsection{Stellar evolutionary isochrones}
From the literature, we can find many public isochrones. However, different models
focus on different stellar evolutionary processes. If one wants to apply those SSP libraries to local galaxies,
which include both old and young, metal-poor and metal-rich stars, we need to select those isochrones that
can cover a wide range of stellar ages, metallicities, and evolutionary phases. We thus select the widely
used Padova2000 \citep{girardi2000, marigo2008} and BaSTI \citep[][]{pietrinferni2004, cordier2007} evolutionary 
isochrones for comparison.

Both the Padova and BaSTI groups track the stellar evolution from the main sequence to thermally pulsating 
asymptotic giant branch states, but their treatments of the thermally pulsating asymptotic giant branch phase 
are very different. A detailed comparison of these two isochrones is shown in the Figure 1 of \cite{CG2010}. 

\subsection{SSP libraries: Galaxev/STELIB and Vazdekis/MILES}
Our analyses on model uncertainties mainly focus on comparing the relative parameter offsets,
$\Delta P = P_{\rm LibA}-P_{\rm LibB}$, where $P_{\rm LibA}$ and $P_{\rm LibB}$ correspond to the 
stellar population parameters derived from the pPXF fitting with two different SSP libraries, respectively.
For each ingredient required for constructing a SSP library, we will check the parameter offsets 
caused by its variation.

To compare the two assumptions for each ingredient (the Salpeter and Chabrier IMFs,
the MILES and STELIB stellar spectral libraries, the BaSTI and Padova stellar evolutionary isochrones),
\cite{bc2003} and \cite{vazdekis2010} provide matched SSP libraries for our tests\footnote{Here we only apply the already
published and widely used SSP libraries for tests.}.
We then select the Galaxev/STELIB and Vazdekis/MILES model based SSP libraries for observational and simulation analyses.
The five SSP libraries listed in Table \ref{ssp_lib} are finally selected with different combinations of the 
three ingredients, by which we can perform our analyses on relative parameter offsets caused by the variation of each ingredient.

\begin{table}
\caption{Five SSP libraries with different combinations of ingredients}
\begin{center}
\begin{tabular}{|c|c|c|c|}
\hline
SSP library    &   model      &     IMF       &   Isochrone \\
\hline
Lib1      &   Vazdekis/MILES   &   SALP    &     P00 \\
\hline
Lib2      &   Vazdekis/MILES   &   CHAB    &     P00 \\
\hline
Lib3      &   Vazdekis/MILES   &   SALP    &     BaSTI \\
\hline
Lib4      &   Galaxev/STELIB      &   SALP    &     P00 \\
\hline
Lib5      &   Galaxev/STELIB      &   CHAB    &     P00 \\
\hline
\end{tabular}
\end{center}
\small
Note: Here the SALP is for Salpeter, the CHAB is for Chabrier, and P00 is for Padova2000.
\label{ssp_lib}
\end{table}

Since the empirical MILES library contains more stars and covers more thorough fundamental parameter space
than STELIB, we also adopt the Vazdekis \citep{vazdekis2010} model with empirical MILES library (Vazdekis/MILES)
to check the model effects of varied IMFs and isochrones. As shown in Table \ref{ssp_lib}, 
Lib1 and Lib2 can be used for checking the parameter offsets caused by the Salpeter 
and Chabrier IMFs, while Lib1 and Lib3 are selected for checking the difference between the Padova and BaSTI isochrones.
To identify the effect of different empirical libraries, we take Lib1 and Lib4 for comparison, where Lib4 is generated
based on the Galaxev \citep{bc2003} evolutionary model with empirical STELIB library (Galaxev/STELIB). In this comparison 
we assume the Padova isochrone and Salpeter IMF as Lib4 for comparing the empirical MILES and STELIB libraries. 
Here we also select Lib5 with the combination of Galaxev/STELIB+Padova isochrone+Chabrier IMF for a double-check of IMF effects. 
Both the Salpeter and Chabrier IMFs assumed in the Vazdekis/MILES and Galaxev/STELIB models have the same stellar mass range: (0.1, 100)$M_{\sun}$.

For both the Lib1 and Lib2 SSP libraries, we select six metallicities ([M/H]=$-1.71$, $-1.31$, $-0.71$, $-0.4$, 0.0, 0.22) and a subset of 
25 logarithmically-spaced equally-sampled ages between 0.06 and 15.85 Gyr. 
For Lib3, which is based on the BaSTI model, the age and metallicity grids are different from that of Lib1 and Lib2. 
Therefore, we select those grids closest to the SSP ages and metallicities selected above, respectively. The metallicity grids are 
[M/H]=$-1.79$, $-1.26$, $-0.66$, $-0.35$, 0.06, 0.26. The age grids are $t/$Gyr=0.06, 0.08, 0.09, 0.1, 0.15, 0.2, 0.25, 0.3, 0.4, 0.5, 0.6, 0.8, 
1.0, 1.25, 1.5, 2.0, 2.5, 3.25, 4.0, 5.0, 6.5, 8.0, 10.0, 12.5, 14.0. For Lib4 and Lib5, their metallicity grids are the same as Lib1 and Lib2, but 
the age grids have small difference. We hence select those age grids closest to that in Lib1 and Lib2 for tests.

\begin{figure*}
\centering
\includegraphics[angle=0.0,scale=0.95,origin=lb]{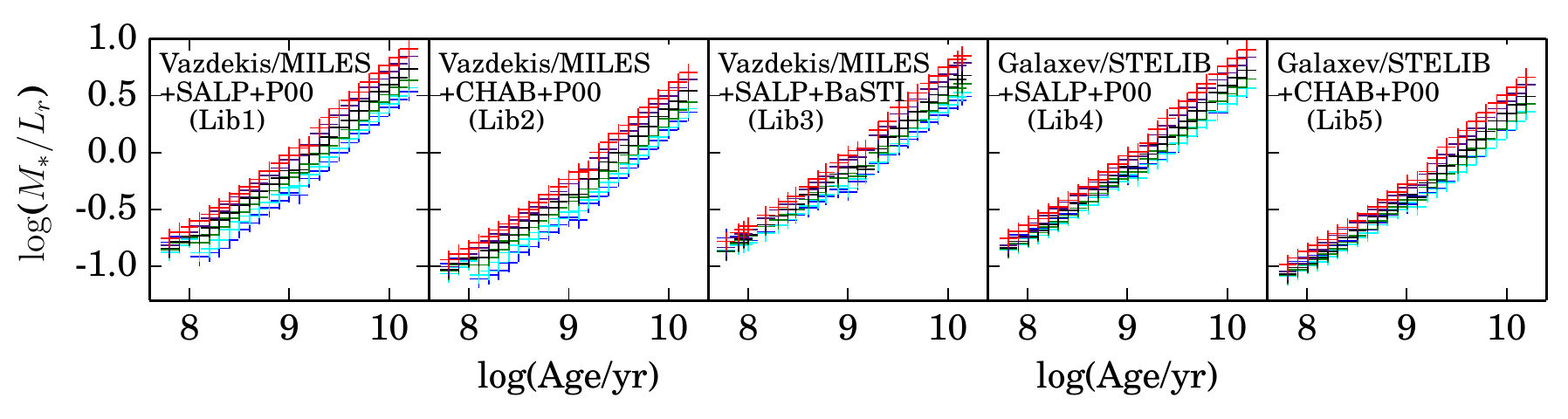}
\caption{The stellar age vs. $M_*/L_r$ in six metallicities of the Lib1 to Lib5 (from left to right panels) 
SSP libraries. In each panel, 
the stellar metallicities are shown in different colors: $\rm [M/H]$= -1.71 (blue), -1.31 (cyan), -0.71 (green), 
-0.4 (black), 0.0 (violet), 0.22 (red).
}
\label{stel_lib}
\end{figure*}

Figure \ref{stel_lib} shows the parameter distributions of 
the stellar age vs. $M_*/L_r$ for the five SSP libraries listed in Table \ref{ssp_lib}.
Different from the middle panel of Figure \ref{imf}, here the models do include the effects 
of mass loss and stellar remnants.
SSPs with the same age but higher metallicities usually have lower mass loss rate and higher
$M_*/L_r$ than that with lower metallicities. However, the $M_*/L_r$ increasing trend of the 
Lib1 and Lib2 libraries are different from the Lib3, Lib4, and Lib5 libraries. 
In the five panels of Figure \ref{stel_lib},
for models with different metallicities, the Lib1 and Lib2 libraries give similar $M_*/L_r$ 
dependences on metallicities at different stellar ages.
While the Lib3 to Lib5 libraries tend to have smaller $M_*/L_r$ differences
among different metallicities
at young ages ($t<10^9$yr), but these differences increase towards older ages ($t>10^9$yr).

As shown in the right panel of Figure \ref{imf}, the variation of $\Delta \log(M_*/L_r)$ along stellar evolution have the similar trends
for both the Vazdekis/MILES (the plus symbols) and 
Galaxev/STELIB (the cross symbols) models
when $t<10^9$ yr or $M_{\rm Upper~ limit}>20 M_{\sun}$ 
(the middle panel of Figure 1).
These variation trends are also consistent for different metallicities until $t\sim 10^{9.5}$yr. 
Then the $M_*/L_r$ offset becomes smaller and those SSPs with poorest metallicity have the smallest
offset. One important thing should be pointed out: although applying the 
same Salpeter and Chabrier IMFs for generating SSP libraries, the $\log(M_*/L_r)$ differences calculated by the 
Vazdekis/MILES model are $\sim 0.05$ dex systematically smaller than those by the Galaxev/STELIB model. 
Even assuming the same stellar evolutionary isochrone and IMF for input, the $M_{*,remn}$ at the same age and metallicity given by the Vazdekis and Galaxev models are different. We have to keep this in mind when 
comparing the Lib1 and Lib4 libraries to study the model offsets caused by the empirical MILES and STELIB 
stellar spectral libraries.

\subsection{MaNGA data analysis}
With the selected pPXF code and five SSP libraries, we can apply them to fit SDSS-IV/MaNGA
spectra and identify parameter offsets caused by different model uncertainties.
Considering current IFU observations have low S/N ($<1$) at large
galaxy radii, we select only spaxels with S/N$>$5 and spatially rebin them
to reach S/N$\sim$60 based on the Voronoi 2D binning method \citep{CC2003}. 
Here the S/N of each spaxel is defined by:
\begin{equation}
	S/N = {\rm median} \left( \frac{F_{\lambda}}{F_{\lambda,err}}\right),~~ \lambda = 5490\angstrom-5510\angstrom.
\end{equation}

For these spatially re-binned spectra, we perform the pPXF fitting by assuming a uniform 
CAL \citep{Calzetti2000} dust reddening curve to correct the intrinsic dust extinction.
This work aims to distinguish the effect of models choices using a state-of-the-art spectral 
fitting code with different SSP libraries and to estimate random uncertainties from ensembles of 
50 Monte-Carlo simulations. For the pPXF fits, we do not use regularisation (we set ``regul=0''). 
This is because, while regularization is useful to reduce the effect of noise on a single fit to 
a single galaxy spectrum, by design it will reduce the scatter in Monte Carlo realization by making 
the fitted weights of all realizations look similar. For this reason, for a proper estimate of 
model uncertainties, regularization should not be used 
\citep[see a similar discussion in Section 3.4 of][]{CappellariEmsellem2004}.
In the spectral fitting, we also mask those emission lines as done in \cite{ge2018}. 
Although the pPXF code can fit emission lines together with the stellar population templates, 
which might improve the fitting quality especially for the Balmer lines, we choose to mask 
the emission lines as this can be applied equally to both the real data and the simulations. 
The simulations using combinations of SSPs based on the assumed log-normal SFH (see Appendix \ref{app_simu} 
for details) are needed to interpret the 
results derived from MaNGA data. Therefore, in this work, we mask emission line regions in both 
the observed and the simulated spectra in exactly the same way. 

When comparing different SSP libraries, the reduced $\chi^2$ is an effective parameter for finding
out their differences. However,
this value depends on the accuracy of flux uncertainty estimation. To avoid the over- or under-estimation
of flux uncertainty, we first use pPXF+Lib1 to fit a spectrum and obtain its best fitting model. 
Then we can derive the spectral fitting residuals by subtracting the best fitting model. 
The root mean square ($\sigma_{\rm resid}$) is hence derived based on the spectral residuals. 
We finally use this $\sigma_{\rm resid}$ to normalize the error spectrum as follows:
\begin{equation}
    F_{\rm err, \lambda} = F_{\rm err, \lambda} \times \sigma_{\rm resid}/median(F_{\rm err}).
\end{equation}
With this normalization, we can 1) keep the shape of the error spectrum along wavelength, which 
reflects the instrumental dependence, and 2) avoid the possibly mis-estimated flux uncertainty.

The uncertainties of those fitted population parameters are estimated based on simulation. 
For each spectrum, we first use the pPXF code and selected SSP library to obtain the best fitting 
model. Then we generate 50 mock spectra based on the best fitting model and the corrected error spectrum
by assuming the observed flux at each wavelength point follows a Gaussian distribution. After performing
the same fitting procedure as done to the observed spectrum, we can derive the parameter uncertainties 
based on the 50 fitted values of each parameter.

Once we obtain the full-spectrum fitting results, we further divide
these spectra into six metallicity bins (as shown in Figure \ref{stel_lib}) based on their luminosity-weighted 
metallicities derived from pPXF+Lib1. In each metallicity bin, we identify each spectrum with SDSS $r-$band 
luminosity-weighted age. With these two steps, we can then perform detailed analyses of parameter offsets 
caused by different model uncertainties.

\section{Results}

The effects of model uncertainties can be reflected by different metrics. The reduced $\chi^2$ 
of full-spectrum fitting shows the fitting quality directly, from which we can compare different 
model assumptions. For those ingredients that can not be differentiated by the reduced $\chi^2$ 
distributions, we analyze their fitting offsets and scatters by a series of comparisons, and 
finally provide a judgment on how suitable they are at fitting different types of galaxies.

\begin{figure}
\centering
\includegraphics[angle=0.0,scale=0.8,origin=lb]{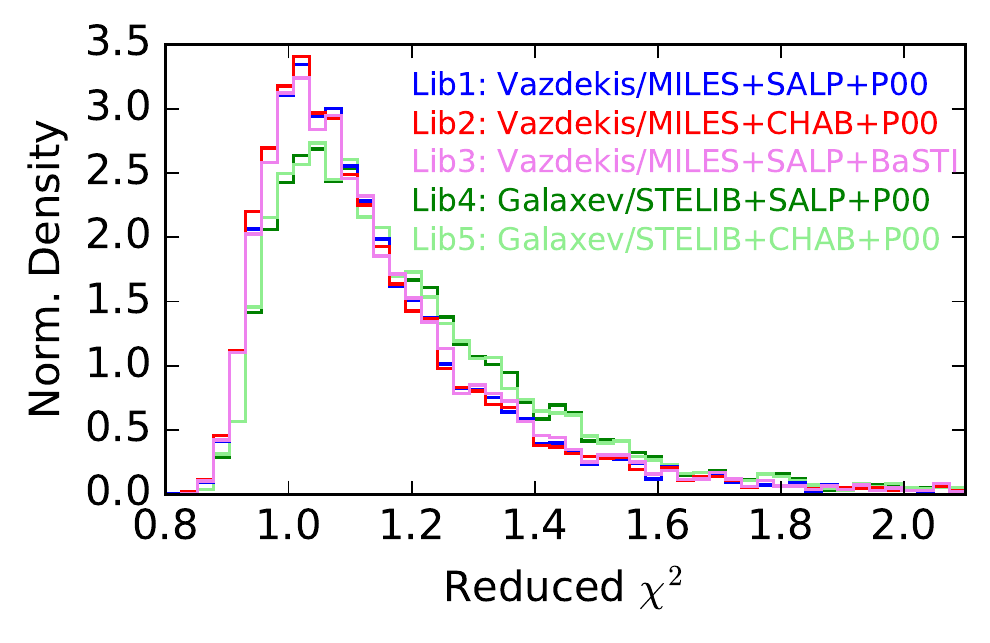}
\caption{
The reduced $\chi^2$ distributions of the MaNGA full-spectrum fitting results based on the five SSP libraries (Lib1$-$Lib5) listed in
Table \ref{ssp_lib}. Distributions derived from Lib1 to Lib3 are based on the Vazdekis/MILES model: 
Lib1 corresponds to Vazdekis/MILES model + Salpeter IMF + Padova2000 isochrone (blue color), 
Lib2 corresponds to Vazdekis/MILES model + Chabrier IMF + Padova2000 isochrone  (red color), 
and Lib3 corresponds to Vazdekis/MILES model + Salpeter IMF + BaSTI isochrone  (violet color).
The reduced $\chi^2$ distributions derived from the Galaxev/STELIB model are shown in green or light green colors.
Here Lib4 defines the combination of Galaxev/STELIB + Salpeter IMF + Padova2000 isochrone (green color) and Lib5 defines 
that of Galaxev/STELIB + Chabrier IMF + Padova2000 isochrone (light green color).
}
\label{chi2}
\end{figure}
\begin{figure*}
\centering
\includegraphics[angle=0.0,scale=0.9,origin=lb]{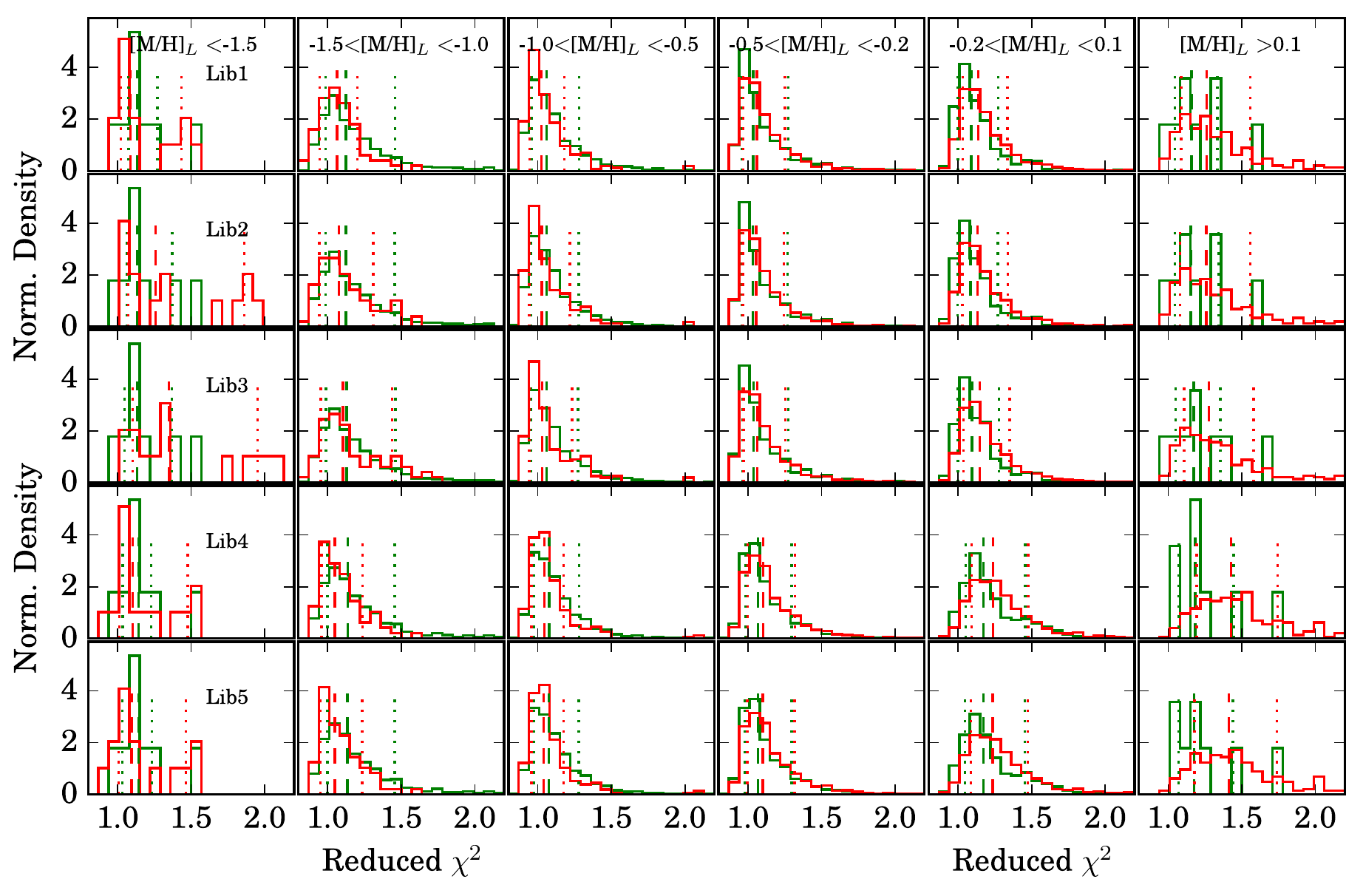}
\caption{
The reduced $\chi^2$ distributions of the MaNGA full-spectrum fitting results based on Lib1 to Lib5 SSP libraries.
From top to bottom rows, we show the reduced $\chi^2$ distributions resulting from the fittings by using Lib1 to Lib5.
To match the input SSP metallicity range, we use the population parameters derived from pPXF+Lib1 to divide the fitted [M/H]$_L$ into six bins: 
$\rm [M/H]<-1.5$ (Column 1), $\rm -1.5<[M/H]<-1.0$ (Column 2), $\rm -1.0<[M/H]<-0.5$ (Column 3), 
$\rm -0.5<[M/H]<-0.2$ (Column 4), $\rm -0.2<[M/H]<0.1$ (Column 5), and $\rm [M/H]>0.1$ (Column 6).
In each panel, the stellar ages derived from pPXF+Lib1 are divided into two bins to describe the corresponding fitting dependences:
$t_L<3$ Gyr (green color), and $t_L>3$ Gyr (red color).
For each age bin, we show the median $\chi^2$ with a vertical dashed line, and show the 16th and 84th percentiles with 
two vertical dotted lines.
}
\label{chi2_bins}
\end{figure*}

\subsection{$\chi^2$ comparisons: the dominant effect of full-spectrum fitting quality}
To compare the fitting quality of different models, we randomly select 100 MaNGA galaxies observed with the 127-fiber bundle, 
which include 26 elliptical and 74 spiral galaxies as classified by the Galaxy Zoo project \citep{Willett2013}.
Those spectra with low S/N ($\rm 5<S/N<60$) are spatially re-binned to S/N$\sim$60
by the Voronoi 2D binning method \citep{CC2003}, which make sure all the spectra we applied for full-spectrum fitting
have sufficiently high S/N. There are $\sim 6000$ spatially re-binned spectra in total, sampled at different galaxy radii. 

When fitting the observed spectra, the measured parameter biases include: 1) algorithm bias, 2) S/N effect, 
and 3) model bias. From our test in \cite{ge2018}, there are almost no algorithm bias caused by the pPXF code at very high S/N (e.g. 300).
At S/N=60, the biases and scatters of age, metallicity and $M_*/L_r$ are all less than 0.1 dex 
(See Appendix \ref{simu_bias}). Based on this, 
we thus apply the pPXF code to fit 
those MaNGA spectra with S/N$\ge$60, and then compare the relative parameter offsets ($\Delta P$) 
derived from different model assumptions.
For those relative offsets larger than 0.1 dex, they should be caused by model uncertainties, 
instead of any algorithm bias or S/N effect.

We start from comparing the reduced $\chi^2$ distribution of full-spectrum fitting based on different combinations of the three ingredients.
From this, we try to identify the dominant ingredient for improving the fitting quality in the current setup of SSP libraries.

Figure \ref{chi2} shows the reduced $\chi^2$ distribution from the fitting of 100 MaNGA galaxies. 
Here the reduced $\chi^2$ is defined as $\chi^2/$DOF, where DOF is the degree of freedom of each spectrum.
Fitting results based on different SSP libraries are divided into two groups:
\begin{itemize}
\item The reduced $\chi^2$ distributions derived from three Vazdekis/MILES-based SSP libraries: 
Lib1 (Vazdekis/MILES+SALP+P00, blue color), Lib2 (Vazdekis/MILES+CHAB+P00, red color), 
and Lib3 (Vazdekis/MILES+SALP+BaSTI, violet color);
\item The reduced $\chi^2$ distributions derived from two Galaxev/STELIB-based SSP libraries:
Lib4 (Galaxev/STELIB+SALP+P00, green color) and Lib5 (Galaxev/STELIB+CHAB+P00, light green color).
\end{itemize}

The five reduced $\chi^2$ distributions shown in Figure \ref{chi2} provide us the following information:
\begin{enumerate}
\item[(1)] The Vazdekis/MILES model based SSP libraries (Lib1--Lib3) provide significant improvement on the fitting quality (colored lines in the left panel) 
compared to Galaxev/STELIB based ones (Lib4 and Lib5). As we can see, the reduced $\chi^2$ distribution with the STELIB library shows a decrease in 
the peak around $\chi^2$/DOF$\sim$1 and an extra tail at large $\chi^2$/DOF$\gtrsim$1.2;
\item[(2)] Different IMF assumptions show little effects to the reduced $\chi^2$ distribution (Lib1 vs. Lib2, and Lib4 vs. Lib5);
\item[(3)] The selection of isochrones also show little effects to the fitting quality (Lib1 vs. Lib3).
\end{enumerate}

From Figure \ref{chi2}, we can only see a significant fitting improvement by Vazdekis/MILES model based libraries, 
but the reason of this improvement requires further clarification. Based on the luminosity-weighted stellar ages and metallicities
derived from pPXF+Lib1, 
we further divide each $\chi^2$ distribution into six metallicity bins and three age bins, 
by which we attempt to identify the detailed fitting quality in each bin and find out the detailed difference caused by
the Lib1 to Lib5 libraries.

In Figure \ref{chi2_bins}, the reduced $\chi^2$ distributions fitted with Lib1 to Lib5 libraries are shown
in the top to bottom rows, respectively. Spectra with poor to rich metallicities are shown from the left to right columns. 
We also divide stellar ages into young ($t_L<3$ Gyr, green color), 
and old ($t_L>3$ Gyr, red color) age bins to track the possible evolution variation.

Those reduced $\chi^2$ distributions obtained from the three Vazdekis/MILES SSP libraries (Lib1 in the first row to Lib3 in the third row)
show no obvious difference at each metallicity and age bin. The results fitted with Galaxev/STELIB SSP libraries
(Lib4 in the fourth row and Lib5 in the fifth row) are similar to that with Vazdekis/MILES ones at ${\rm [M/H]}_L<-0.2$ for both the young and old age bins.
In the metal-rich cases (${\rm [M/H]}_L>-0.2$), the reduced $\chi^2$ from Lib4 and Lib5 are systematically larger than that from Lib1 to Lib3, which is 
reflected by median values of the two age bins (vertical dashed line). These median values present the largest differences ($\Delta \chi^2/{\rm DOF} \sim $0.2) for those spectra with $t_L>3$ Gyr
at the largest metallicity bin (${\rm [M/H]}_L>0.1$), and these differences are the same for both Lib4 and Lib5 reduced results. The green histograms in the ${\rm [M/H]}_L>0.1$ bin only include 8 spectra with stellar
ages $1~ {\rm Gyr}<t_L<3$ Gyr. Although their median values are similar, we can not derive a robust statistical conclusion.

\begin{figure*}
\centering
\includegraphics[angle=0.0,scale=0.8,origin=lb]{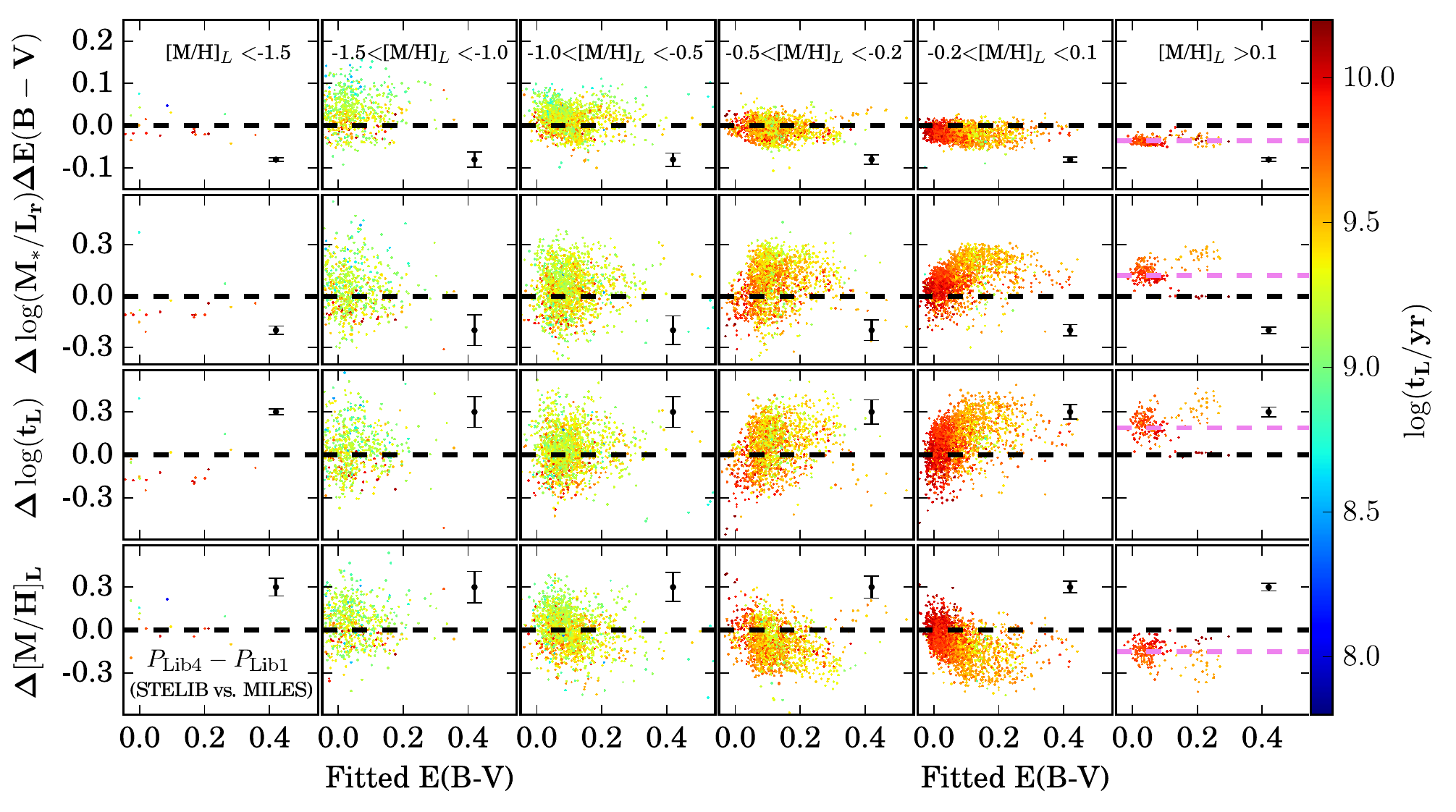}
\caption{
The relative parameter offsets ($\Delta P = P_{\rm Lib4}-P_{\rm Lib1}$) between the spectral fitting 
results of Galaxev/STELIB and Vazdekis/MILES models derived from MaNGA data at S/N=60.
For this comparison, Lib1 and Lib4 libraries are generated with the same Salpeter IMF and Padova2000 isochrones. 
The six metallicity bins from the left to right columns are the same as shown in Figure \ref{chi2_bins}.
Each spectrum is plotted as a point colored by its luminosity-weighted age.
Blue to red colors represent the stellar age ranging from 0.063 to 15 Gyr.
The relative offsets of the four parameters ($\Delta E(B-V)$,
$\Delta \log(M_*/L_r)$, $\Delta \log t_L$, and $[M/L]_L$) are shown from top to bottom.
The zero-offset line of each parameter in each panel is labeled as the horizontal dashed line.
The violet line indicates the median offset. In each panel, the black error bar shows
the median uncertainty of the parameter offsets of those colored data points.
}
\label{miles_bc03_fit}
\end{figure*}
\begin{figure*}
\centering
\includegraphics[angle=0.0,scale=0.66,origin=lb]{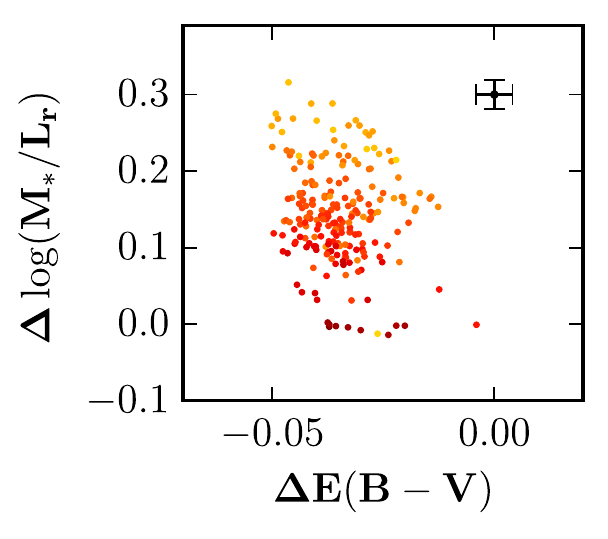}
\includegraphics[angle=0.0,scale=0.66,origin=lb]{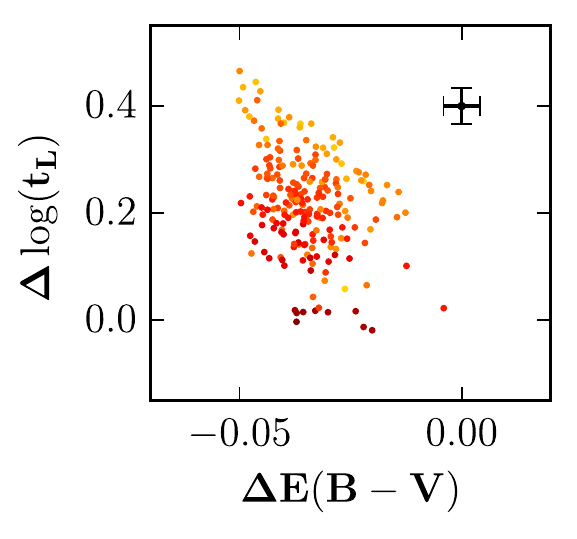}
\includegraphics[angle=0.0,scale=0.66,origin=lb]{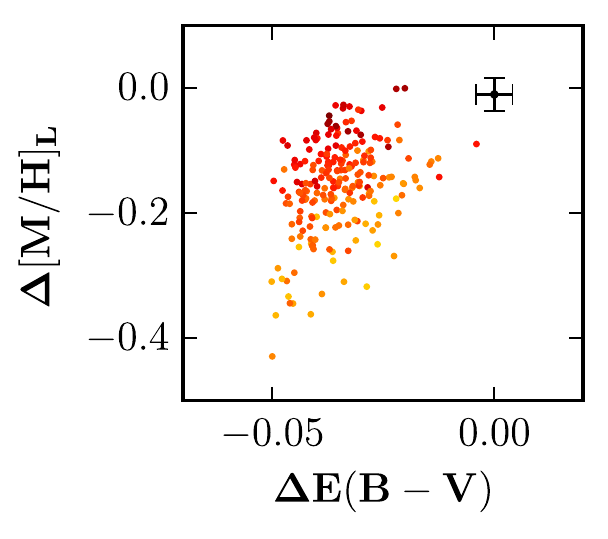}
\includegraphics[angle=0.0,scale=0.66,origin=lb]{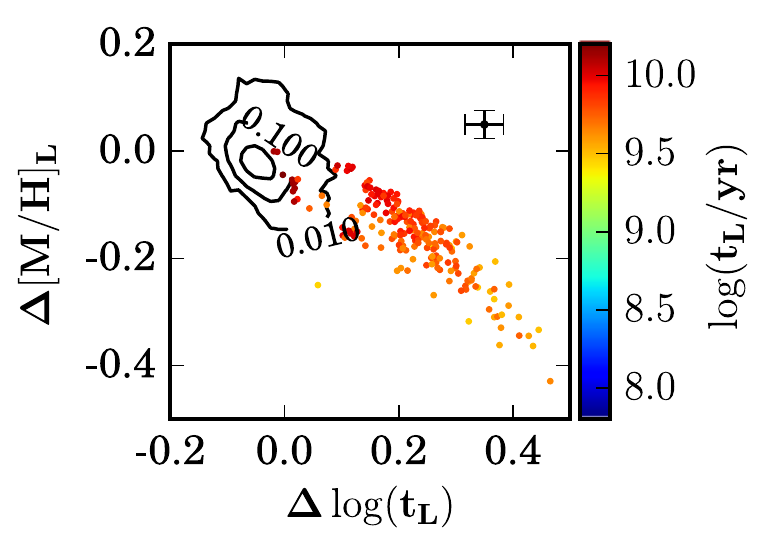}
\caption{Relations between different parameter offsets. The $\Delta$E(B-V) vs. $\Delta \log(M_*/L_r)$,
$\Delta$E(B-V) vs. $\Delta \log(t_L)$, $\Delta$E(B-V) vs. $\Delta {\rm [M/H]}_L$, and $\Delta \log(t_L)$
vs. $\Delta {\rm [M/H]}_L$ plots are shown from the left to right panels. In the right panel, we also
plot the contours from simulation, which show the effect of age-metallicity degeneracy when performing 
the pPXF fitting at S/N=60. Each spectrum is plotted as a point colored by its luminosity-weighted age.
The black error bar in each panel shows the median uncertainty of the parameter offsets.
}
\label{dpara_corrs}
\end{figure*}
\begin{figure*}
\centering
\includegraphics[angle=0.0,scale=0.9,origin=lb]{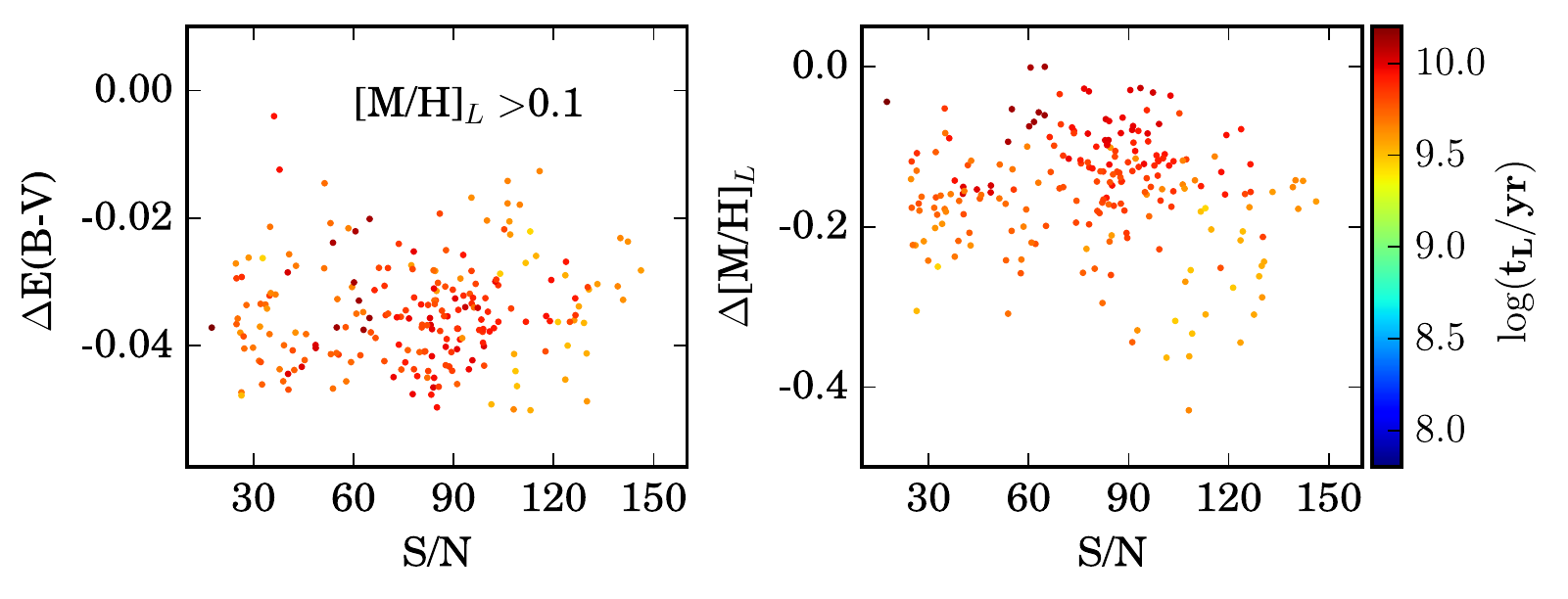}
\caption{Variations of $\Delta$E(B-V) (left panel) and $\Delta$[M/H]$_L$ (right panel) with S/N for the [M/H]$_L>0.1$ bin. The points are colored the same as shown in
Figure \ref{dpara_corrs}.}
\label{snr_comp}
\end{figure*}
\begin{figure*}
\centering
\includegraphics[angle=0.0,scale=0.85,origin=lb]{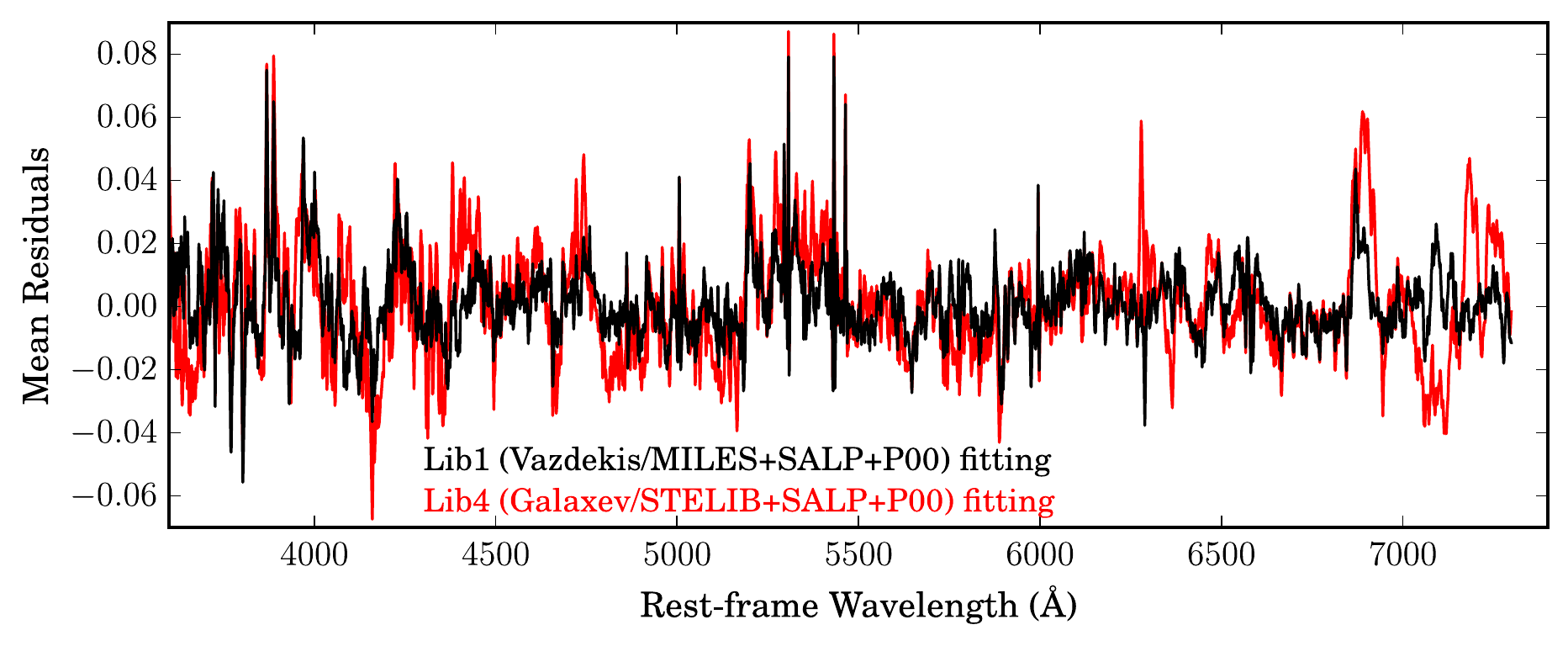}
\caption{The mean fitting residuals of those spectra fitted by the Lib1 (in black color) and Lib4 
(in red color) libraries. To check whether the fitting differences are caused by the relative chemical 
abundances, here we only stack those spectra with $\Delta (\chi^2/{\rm DOF}) > 0.2$, which is the reduced
$\chi^2$ difference derived from the Lib1 and Lib4 libraries.
}
\label{fit_resid}
\end{figure*}

The above results might be explained as follows: 1) The STELIB library lacks metal-rich stars, 
but the MILES library includes sufficient such stars. SSP templates with super-solar metallicities 
are not well calibrated for Lib4 and Lib5 libraries, leading to larger reduced $\chi^2$ than those 
derived from Lib1, Lib2, and Lib3 libraries. 2) The difference between Salpeter and Chabrier IMFs 
mainly happens to those low mass stars ($<1M_{\sun}$), which cause large mass difference but only 
have small light contribution; 3) The isochrones given by Padova2000 and BaSTI models are the same 
in most cases, their difference cannot be differentiated easily in the reduced $\chi^2$ distributions.

Based on the above analyses of the reduced $\chi^2$ distributions, we can only see some of the effects 
on how the choice of ingredients affect the fitting quality.
How these stellar population parameters are biased by different choices of the three ingredients 
(IMF, isochrones and empirical library) require further investigation.

\subsection{Defining the relative parameter offset}
Since the algorithm caused parameter biases and scatters of the pPXF code are all smaller than 0.1 dex 
at S/N$\ge60$ (Appendix \ref{simu_bias}), we can confirm the effects of model uncertainties if the 
corresponding offsets ($\Delta \log(M_*/L_r)$, $\Delta \log(t_L)$, and $\Delta$[M/H]$_L$) are larger 
than 0.1 dex. The dust extinction can also be corrected well by pPXF, hence the measured E(B-V) can 
be confirmed as a offset if the resulting $\Delta$E(B-V) is larger than 0.05 magnitude.

Here we focus on the relative offset of four parameters: E(B-V), $r$-band $M_*/L_r$, $r$-band 
luminosity weighted age ($\log(t_L/\rm yr)$), and $r$-band luminosity weighted [M/H]$_L$ defined 
as follows:
\begin{equation}
M_*/L_r=\frac{\Sigma f_{M,i}}{\Sigma f_{M,i}/(M_*/L_r)_i},
\end{equation}
\begin{equation}
\log(t_L)=\Sigma f_{L,i}\times\log(t_i/yr),
\end{equation}
\begin{equation}
{\rm [M/H]}_L=\Sigma f_{L,i}\times{\rm [M/H]}_i,
\end{equation}
where $f_{M,i}$, $f_{L,i}$, $(M_*/L_r)_i$, $\log(t_i/yr)$, and $[M/H]_i$ correspond to the fitted mass fraction, 
the fitted luminosity fraction, the $r$-band mass-to-light ratio, age, and metallicity of the $i$-th SSP, respectively.
The parameter offsets for each spectral fitting are then calculated by:
\begin{equation}
\Delta {\rm E(B-V)} =  \frac{1}{N} \sum_i \left[{\rm E(B-V)}_{i}^{P_1}-{\rm E(B-V)}_{i}^{P_2}\right],
\end{equation}
\begin{equation}
\Delta \log (M_*/L_r) =  \frac{1}{N} \sum_i \left[\log(M_*/L_r)_{i}^{P_1}-\log(M_*/L_r)_{i}^{P_2}\right],
\end{equation}
\begin{equation}
\Delta \log t_L= \frac{1}{N} \sum_i \left(\log t_{L,i}^{P_1}-\log t_{L, i}^{P_2}\right),
\end{equation}
\begin{equation}
\Delta{\rm [M/H]}_L= \frac{1}{N} \sum_i \left({\rm [M/H]}_{L,i}^{P_1}-{\rm [M/H]}_{L,i}^{P_2}\right).
\end{equation}
where $P_1$ and $P_2$ are the population parameters derived from the two selected SSP libraries, 
respectively.

\subsection{Empirical stellar spectral library effects}
In Figures \ref{chi2} and \ref{chi2_bins}, the lack of metal-rich stars can affect the fitting
quality to those high metallicity spectra, but the selected IMF and isochrones have little effect
to the fitting results. Here we compare the fitting results derived from fittings with Lib1 and 
Lib4 to quantify their relative parameter offsets ($\Delta P = P_{\rm Lib4}-P_{\rm Lib1}$).

Figure \ref{miles_bc03_fit} shows the relative parameter offsets caused by Lib1 and Lib4 by
applying pPXF and corresponding libraries to MaNGA spectra at S/N=60. 
In the six metallicity bins, the biggest difference is shown in the most metal-rich bin.
Due to the lack of metal-rich stars in the STELIB library, those super-solar SSP templates are 
actually constructed based on relatively low metallicity stars, which caused an under-estimation of
the metallicity for those high metallicity spectra.

Based on the current MaNGA galaxy sample, the averaged offsets of the four parameters in the 
largest metallicity bin ([M/H]$_L>0.1$) are shown in violet dashed lines: 
$\Delta \rm E(B-V)=-0.03\pm0.01$ mag, $\Delta \log(M_*/L_r)=0.12\pm0.07$ dex, 
$\Delta \log(t_L)=0.19\pm0.09$ dex, and $\Delta$[M/H]$_L=-0.15\pm0.09$ dex.

For the solar-like metallicity bin ($-0.2<{\rm [M/H]}_L<0.1$), those spectra with age around 5 Gyr or younger
have similar trends as shown in the [M/H]$_L>0.1$ bin. For those spectra with $t_L \sim 10$ Gyr, the
systematic offsets almost disappear. 

For the left four bins with [M/H]$_L<-0.2$, there are no clear systematic offsets, and the parameter 
differences are dominated by scatters.

To explore the dominant reason of those parameter offsets, we first check the relations between 
these parameter offsets in Figure \ref{dpara_corrs}. For those spectra with [M/H]$_L>0.1$, spectral fittings
with the Lib4 library can introduce negative offsets of dust extinction relative to that with the Lib1 library. 
These negative offsets of E(B-V) correspond to positive offsets of $M_*/L_r$ (left panel) and stellar age 
(middle left panel), and negative offsets of [M/H]$_L$ (middle right panel). In the right panel,
we further clarify that the age and metallicity offsets are not caused by the age-metallicity degeneracy.
The effect of the age-metallicity degeneracy when performing pPXF fitting at S/N=60 is shown by the black contours.
Most of the data points in the right panel are outside the contours, which means that 
the parameter offsets $P_{\rm Lib4}-P_{\rm Lib1}$ are not caused by the age-metallicity degeneracy.

Considering that the S/N of our observed spectra are not uniformly 60, we then plot the dust extinction
and metallicity offsets versus the spectral S/N's in Figure \ref{snr_comp}. There is no obvious trend
between $\Delta$E(B-V) or $\Delta$[M/H]$_L$ offsets and S/N, which means that these parameter offsets 
are also not caused by the S/N effect.

Without the dominant reason from the age-metallicity degeneracy and S/N effects, the paremeter offsets of 
those high metallicity spectra are most probably caused by the differences between the empirical STELIB and
MILES libraries, or code improvement by the Vazdekis model. Since the STELIB library not only lacks the 
metal-rich stars, but also has other differences
compared to MILES, e.g. not corrected from the telluric absorption, we then stack those spectra with 
$\Delta \chi^2/$DOF$>0.2$ to identify whether these $\chi^2$ differences are caused by the continuum or
relative chemical abundance. From Figure \ref{fit_resid} we can see that the stacked fitting residuals 
obtained from the Lib1 (black line) and Lib4 (red line) libraries do not suffer a continuum correction problem.
Their main differences happen to those absorption-line features, which indicates that the parameter offsets
are most probably caused by differences of the relative chemical abundances. Here we cannot exclude 
the possibility of code improvement by the Vazdekis model without detailed comparisons of the two models, 
which is out of the scope of this paper.

With the above reduced $\chi^2$ and parameter offsets analyses, we find that the Vazdekis/MILES model 
can provide better spectral fitting quality than the Galaxev/STELIB model for [M/H]$_L>-0.2$ cases. 
Thus, we take the Vazdekis/MILES model for analyzing the effects of IMF variation and stellar evolution isochrones.

\subsection{IMF variation effects}\label{sec_imfvar}
Considering that the empirical MILES library not only has more metal-rich stars, but also covers more 
complete fundamental parameter space than the STELIB library. We study the IMF variation effects by 
applying Lib1 and Lib2 to the data to obtain the relative parameter offset 
($\Delta P = P_{\rm Lib2}-P_{\rm Lib1}$), 
focusing only on the difference between Salpeter and Chabrier IMFs, 
and keeping the same Padova2000 isochrone and empirical MILES library. 

\begin{figure*}
\centering
\includegraphics[angle=0.0,scale=0.8,origin=lb]{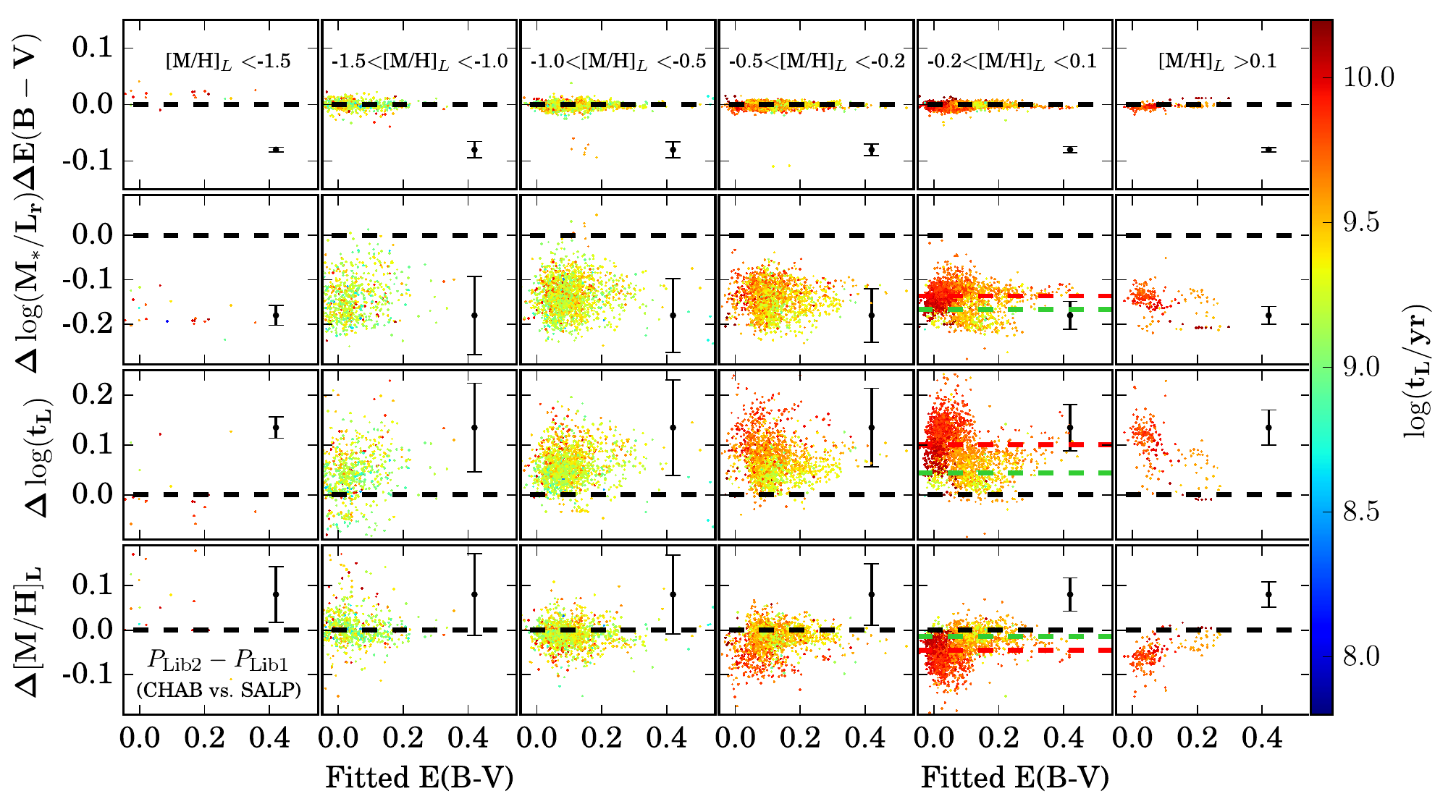}\\
\includegraphics[angle=0.0,scale=0.8,origin=lb]{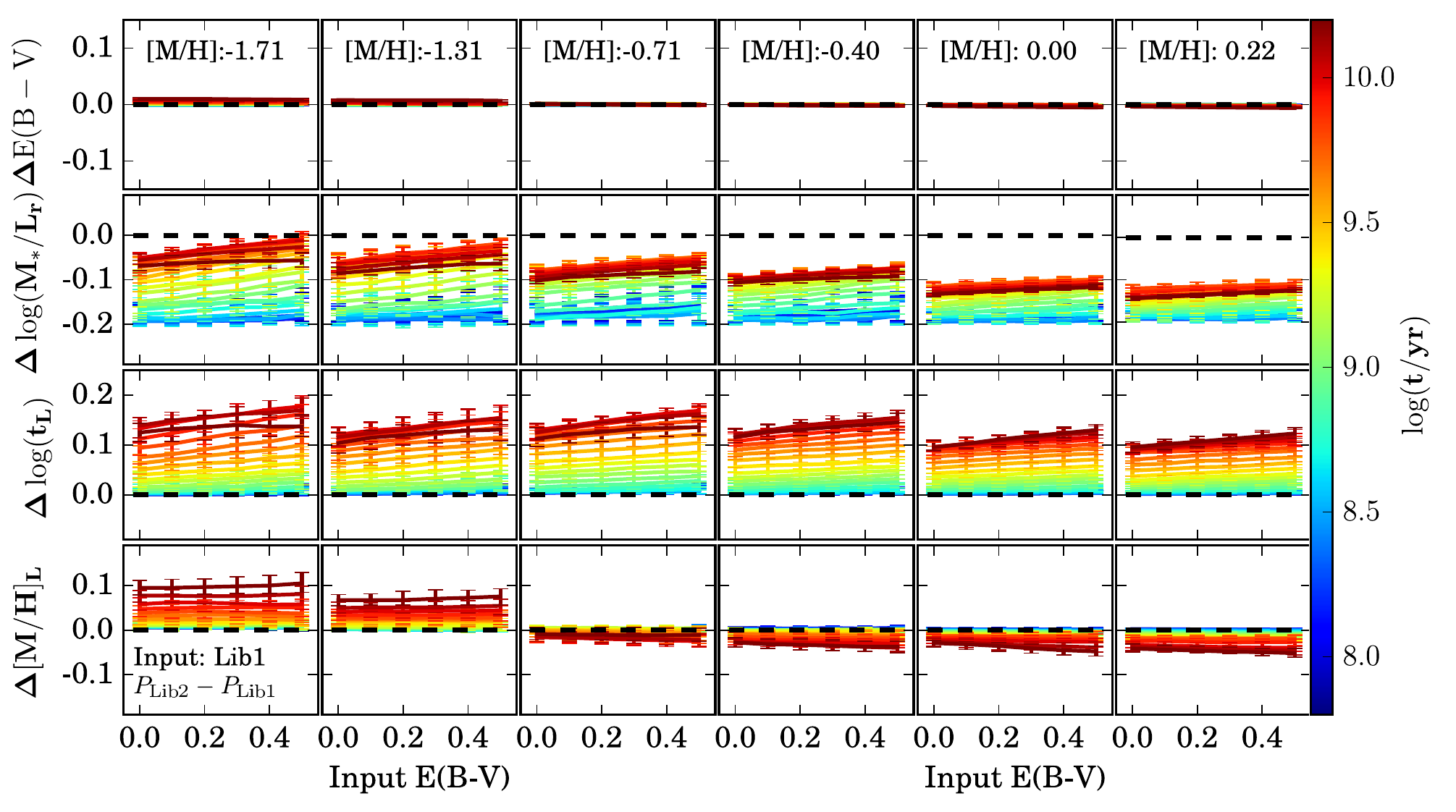}
\caption{
$Top:$ The relative parameter offsets ($\Delta P = P_{\rm Lib2}-P_{\rm Lib1}$) between the Chabrier 
and Salpeter IMFs in the case of MaNGA spectral fitting at S/N=60.
In this comparison, Lib1 and Lib2 have the same Vazdekis/MILES model and Padova2000 isochrones.
Lines, points, error bars, and colors are the same as described in Figure \ref{miles_bc03_fit}.
In the fifth column, which contains the largest number of data points, we plot the red and green dashed lines to 
show the median offsets ($\Delta \log(M_*/L_r)$, $\Delta \log t_L$, and $\Delta$[M/H]$_L$) of those 
old ($t_L>3$ Gyr) and young ($t_L<3$ Gyr) spectra, respectively.
$Bottom:$ We generate mock spectra based on Lib1, and then perform the fitting with pPXF+Lib1 and pPXF+Lib2.
The relative parameter offsets ($P_{\rm Lib2}-P_{\rm Lib1}$) in the simulation can be compared with that shown in the top figure.
From left to right, we show the six metallicity bins ([M/H]=$-1.71$, $-1.31$, $-0.71$, $-0.4$, 0.0, 0.22) bins based 
on the peak [M/H] of each SFH. Blue to red colors represent
the stellar age ranging from 0.063 to 15 Gyr. The offsets in the four parameters ($\Delta$E(B-V),
$\Delta \log(M_*/L_r)$, $\Delta \log t_L$, and $\Delta$[M/H]$_L$) are shown from top to bottom.
The zero-offset line of each parameter in each panel is labeled as the horizontal dashed line. 
The error bars indicate the 16th and 84th percentiles.
}
\label{miles_imf}
\end{figure*}

In the top panel of Figure \ref{miles_imf}, we show the difference in MaNGA spectral fitting results between pPXF+Lib1 and pPXF+Lib2. The parameter offsets caused by IMF variation behave as follows:
\begin{itemize}
\item The IMF variation causes no significant E(B-V) fitting offset;
\item The $\Delta \log(M_*/L_r)$ ranging from $-0.2$ to $-0.1$ dex, and the offset values correlate 
with stellar ages;
\item Spectra with old ages (red dashed lines) have larger $\log(t_L/\rm yr)$ offsets than those younger 
ones (green dashed lines);
\item The ${\rm [M/H]}_L$ show increased offsets for those spectra with older ages when ${\rm [M/H]}_L>-0.5$.
\end{itemize}

To interpret these relative offsets caused by the Salpeter and Chabrier IMFs, we use simulations to 
find out the main reason. According to the method shown in Appendix A, we generate mock spectra
based on Lib1, then fit these mock spectra using pPXF+Lib1 and pPXF+Lib2, respectively. By analyzing the 
relative parameter offsets calculated from $P_{\rm Lib2}-P_{\rm Lib1}$ (bottom panels of Figure \ref{miles_imf}), 
we can check the IMF variation effects and interpret how and why these population parameters are offset. 
Before comparing with observation, we first summarize the IMF effects in our simulation:
\begin{itemize}
\item There is no offset for the dust extinction correction in all cases.
\item When fitting a spectra with Chabrier IMF, the $M_*/L_r$ will be under-estimated by $\sim 0.2$ dex 
at the young ages ($t_L=10^8$yr). With increased spectral age and poorer metallicity, the $M_*/L_r$ differences decrease.
\item The age offsets increase from zero for the youngest spectra to a maximum for the oldest spectra.
The largest age offsets decrease from metal-poor ($\sim 0.15$ dex) to metal-rich ($\sim 0.1$ dex) bins.
\item The metallicity offsets appear, and the values vary from positive offsets at the smallest metallicity to negative offsets 
at the largest metallicity. These offsets also increase with stellar ages.
\end{itemize}
These $M_*/L_r$, age, and metallicity offsets in the simulation are significant, and the offset
trends described above are not be affected by
the fitting uncertainty, which is reflected by the scatters shown in the bottom panel of Figure \ref{miles_imf}.

With the simulation results, now we can derive the IMF effects on full-spectrum fitting results, which have been shown 
partly by the observation. Based on the simulation, we can also identify how the input SFH can be recovered by
spectral fitting, and the dominant reason of parameter offsets.

\begin{figure*}
\centering
\includegraphics[angle=0.0,scale=0.6,origin=lb]{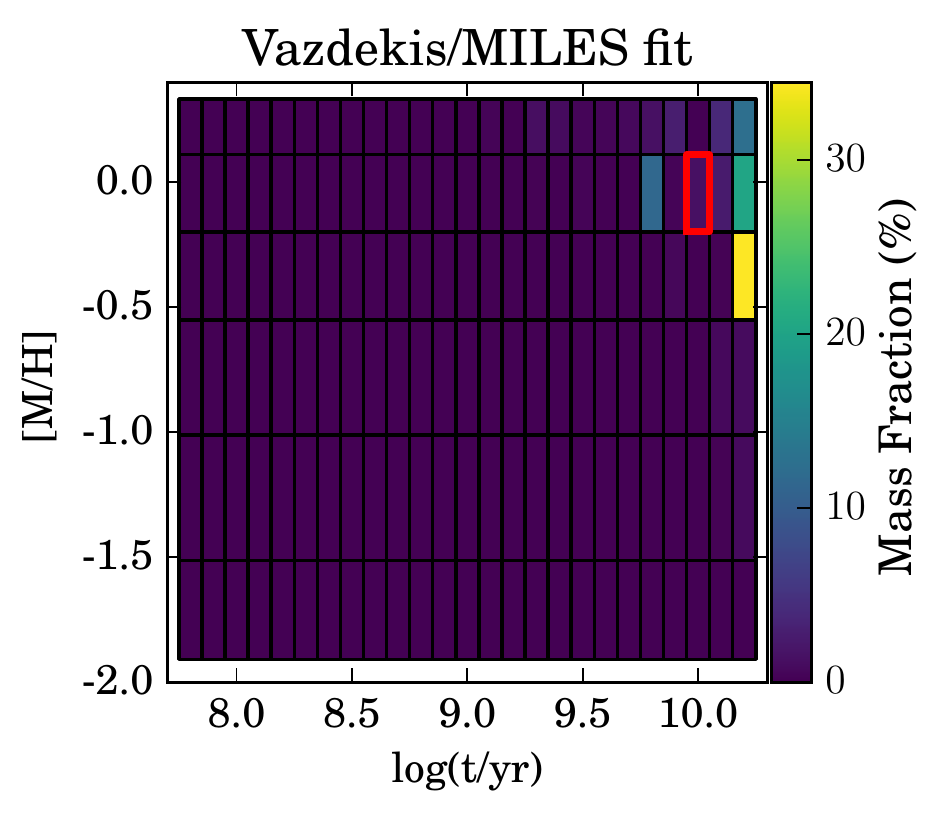}
\includegraphics[angle=0.0,scale=0.6,origin=lb]{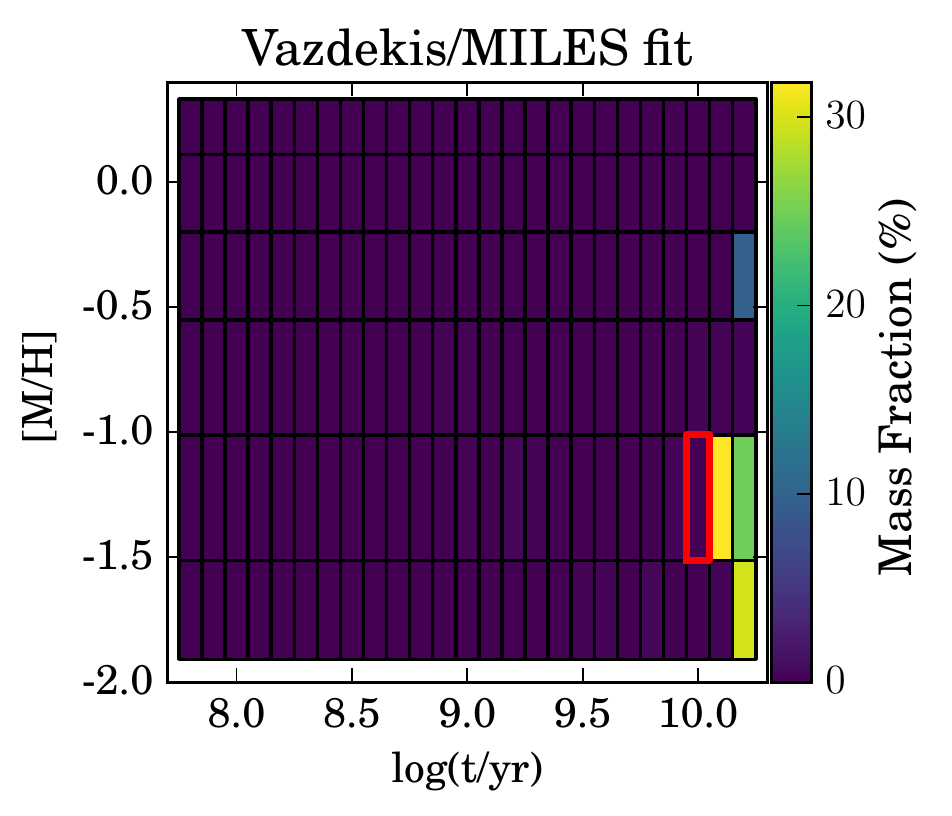}
\includegraphics[angle=0.0,scale=0.6,origin=lb]{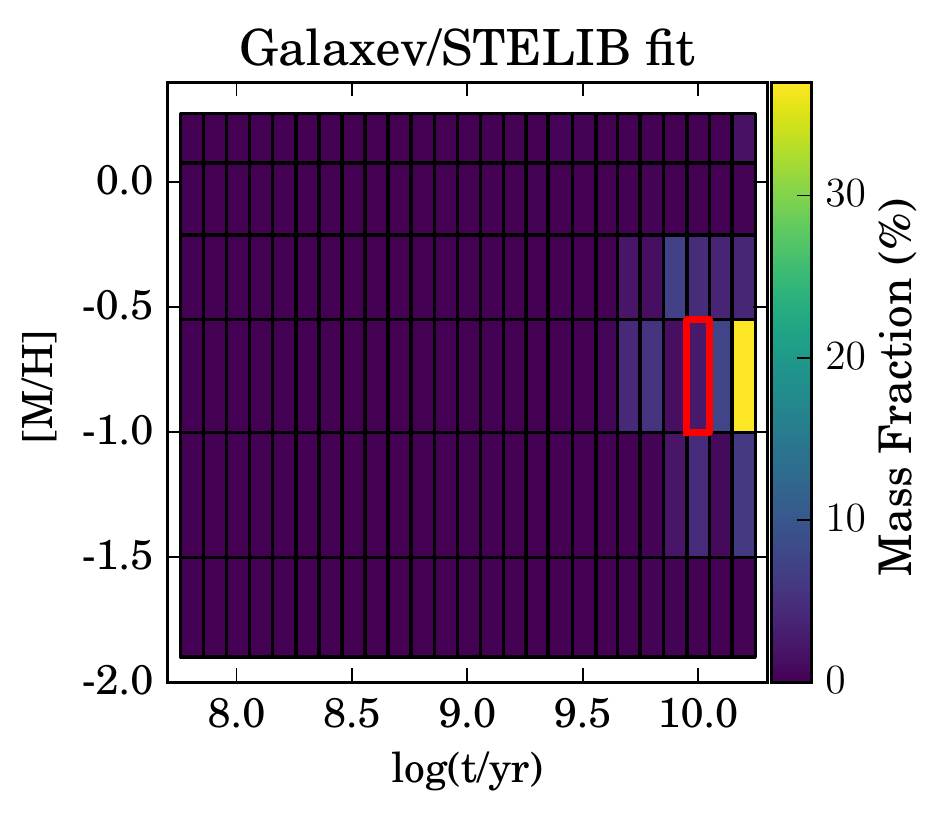}
\caption{
Recovered SFH based on the pPXF fitting to the mock spectra generated with Lib1 (left and middle panels) 
and Lib4 (right panel), but fitted by using Lib2 and Lib5, respectively. The SFH shown in each panel is the averaged result 
calculated from 50 simulations. The peak of the input age and metallicity distributions of each 
mock SFH is labeled by the red box in each panel. From left to right, the peak
ages and metallicities are: [10.0, 0.0], [10.0, $-1.31$], and [10.0, $-0.71$]. As shown in Figure \ref{miles_imf}, spectra
with older ages have larger parameter offsets. Therefore, we take the three spectra with old age as examples.
}
\label{imf_examples}
\end{figure*}

For those spectra with [M/H]$_L>-1.0$, the over-estimated stellar ages and under-estimated metallicities are 
mainly caused by two effects: 1) the intrinsic age-metallicity degeneracy, and 
2) SSP template difference caused by different IMFs at the same age and metallicity. 

Because an SSP template with a younger age and a higher metallicity is similar to the one with an older age 
and a lower metallicity, the positive offset in stellar age propagates to a negative 
offset in metallicity.

At young stellar ages, high mass stars
still exist and those low mass stars contribute very little to the total light. 
As the SSP ages increase and more high mass stars explode, the light fraction contributed by low mass stars 
will increase.  Since the Salpeter IMF contains more low mass stars than the Chabrier IMF, a given age 
SSP generated with the Salpeter IMF would have a similar low-mass-star light fraction as that of the 
Chabrier IMF based SSP with an older age. Therefore, fitting a spectrum with the Chabrier IMF would result 
in an older age than that with the Salpeter IMF. The resulted parameter offsets increase with stellar ages.

In the right four columns of Figure \ref{miles_imf}, 
the first effect explains why there are $\Delta \log(t_L)$ and $\Delta$[M/H]$_L$ offsets,
the second effect explains why the age and metallicity offsets increase with older ages.

The recovered SFHs also provide a clear reflection of these two effects.
Here we take old spectra ($t_L=10$ Gyr) as examples, as the age and metallicity offsets shown in Figure \ref{miles_imf} 
increase with stellar ages.
In the left panel of Figure \ref{imf_examples}, the recovered SFH is dominated by those SSP templates with older ages and lower metallicities.

For the results shown in the left two columns of Figure \ref{miles_imf}, 
the metallicities are measured with positive offsets. This should be mainly due to the limitation of 
model grid, as there are no SSPs with smaller metallicity for fitting. This can be verified
by the middle panel of Figure \ref{imf_examples}, which shows an SSP template at the smallest metallicity with the oldest
age contributing a high mass fraction. Since there is no SSP with a smaller metallicity, the 
fitting finds an SSP with higher metallicity ([M/H=$-0.4$]) to match the input spectrum.

\begin{figure*}
\centering
\includegraphics[angle=0.0,scale=0.8,origin=lb]{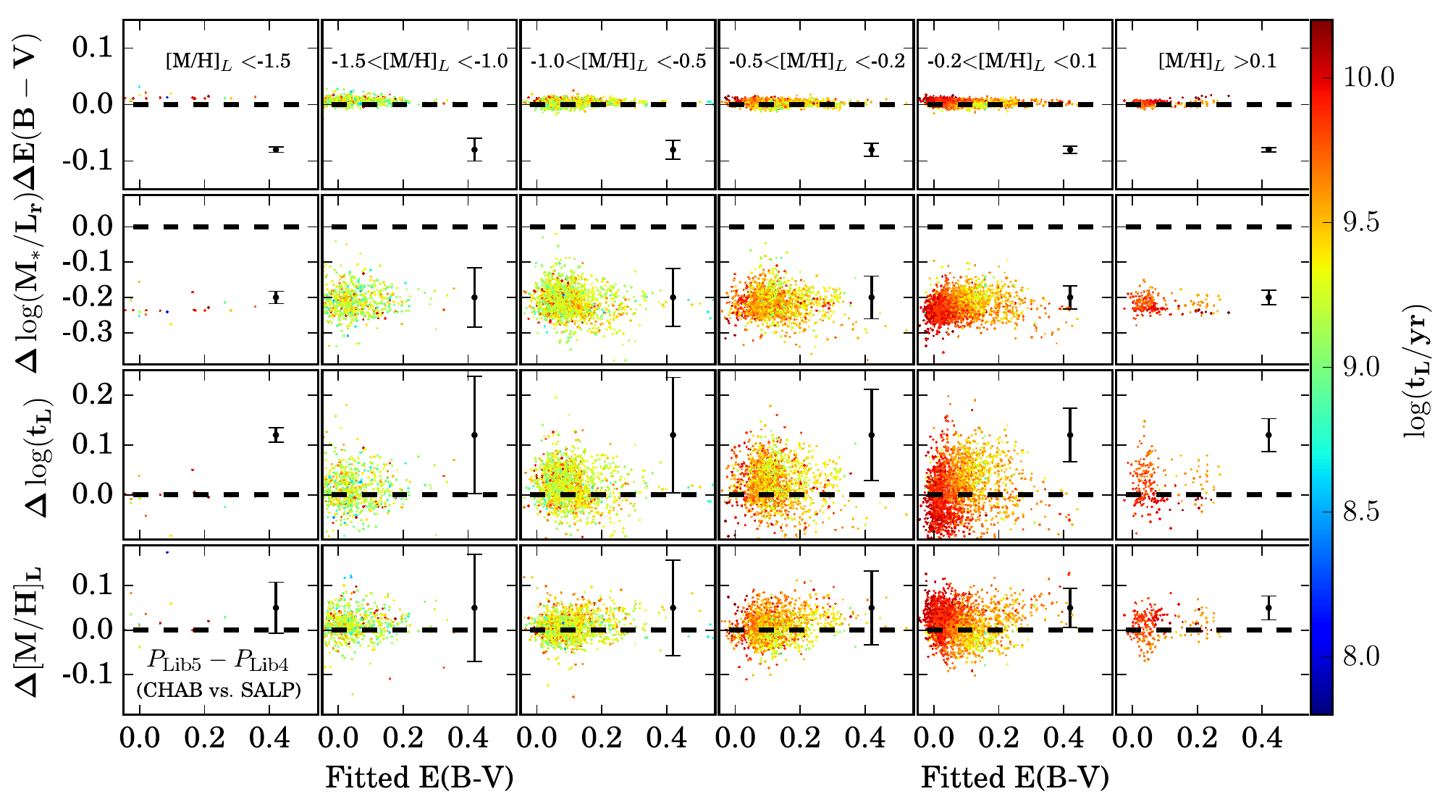}\\
\includegraphics[angle=0.0,scale=0.8,origin=lb]{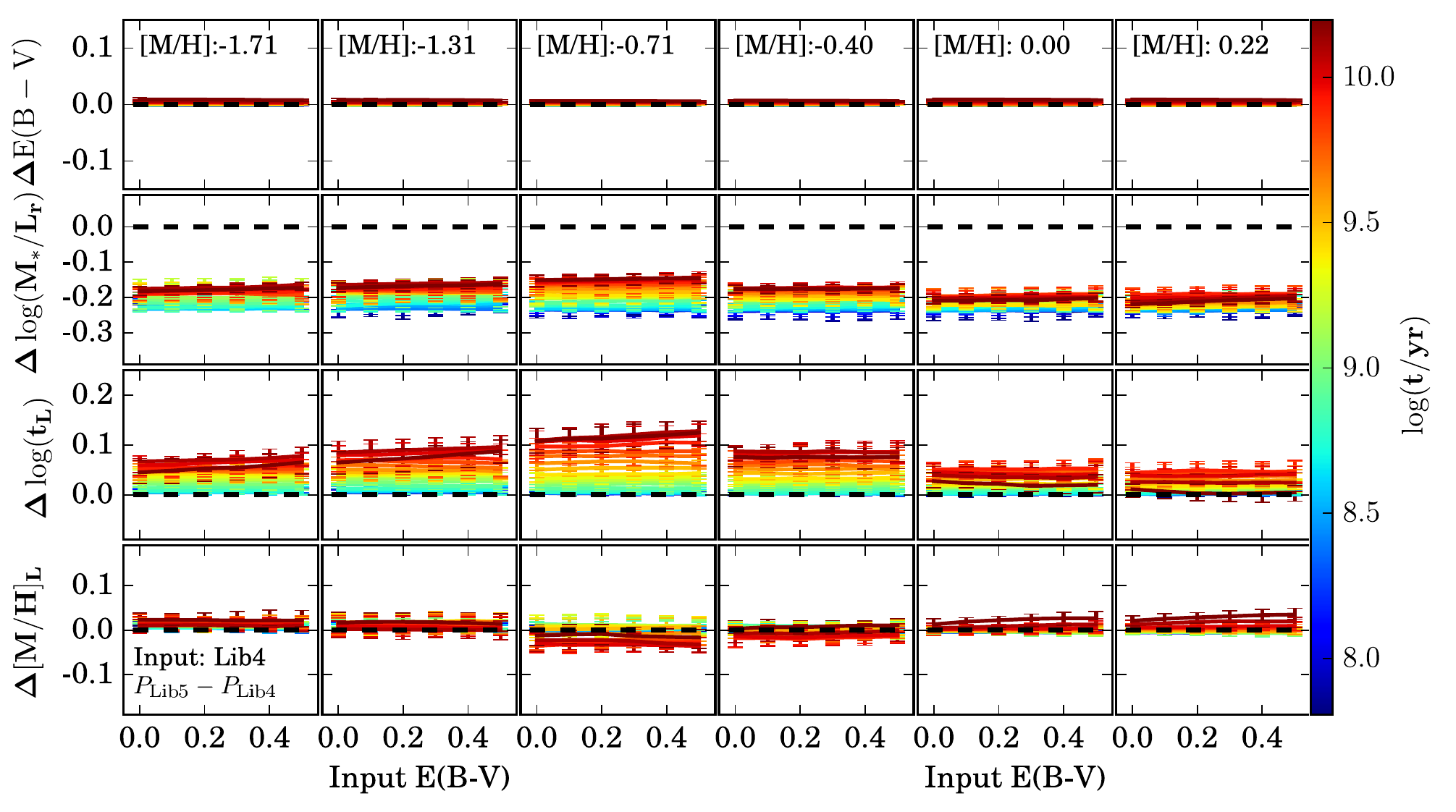}
\caption{
$Top:$ The relative parameter offsets ($P_{\rm Lib5}-P_{\rm Lib4}$) between the Chabrier and Salpeter IMFs
in the case of MaNGA spectral fitting at S/N=60.
In this comparison, Lib4 and Lib5 have the same Galaxev/STELIB model and Padova2000 isochrones.
Lines, points, error bars, and colors are the same as described in Figure \ref{miles_bc03_fit}.
$Bottom:$ With the mock spectra generated based on Lib4, the spectral fittings are performed by pPXF+Lib4 and pPXF+Lib5.
The relative parameter offsets ($P_{\rm Lib5}-P_{\rm Lib4}$) in the simulation can be compared with 
those shown in the top figure. Lines, points, and colors are the same as described in Figure \ref{miles_imf}.
}
\label{bc03_imf}
\end{figure*}

The age-metallicity degeneracy should be affected by the fundamental parameter coverage of the empirical stellar 
spectral library. If using a library with fewer stars, the difference in spectral features between those SSP templates with 
similar ages and metallicities should be more obvious, which means the age-metallicity degeneracy is weak. Therefore, 
if we use the Galaxev/STELIB-based SSP libraries to check the IMF variation effects, the age-metallicity degeneracy should 
affect the parameter offsets less than using the Vazdekis/MILES-based SSP libraries.

Figure \ref{bc03_imf} shows the comparisons of IMF variation effects ($\Delta P = P_{\rm Lib5}-P_{\rm Lib4}$) for 
both observational spectral ($top$) and mock 
spectral ($bottom$) fitting with the pPXF code. Different IMFs still cause no offset to the dust extinction measurements. 
For the Galaxev/STELIB model, the $M_*/L_r$ derived from the Chabrier IMF are systematically lower by a $\sim0.2$ dex 
than that from the Salpeter IMF for both the simulation and observation tests. By comparing Figures \ref{miles_imf}
and \ref{bc03_imf}, we find that there are smaller age and metallicity offsets when apply
the Galaxev/STELIB model based libraries for IMF variation tests. 

From the recovered SFH shown in the right panel of Figure \ref{imf_examples}, the best spectral fitting catches older SSPs with 
$\log(t/\rm yr)$=10.2 rather than $\log(t/\rm yr)$=10.0 as the peak age, which finally causes a systematic $\sim 0.1$ dex age offset.
This is mainly caused by the SSP template difference effect. 
Considering that the empirical STELIB library only contains $\sim 197$ spectra for calibration, the resolution in the metallicity space may
be not sufficiently high. The age-metallicity degeneracy effect
is then not obvious in the application of the Galaxev/STELIB model based libraries.

With the understanding of empirical library and IMF effects, we then focus on the comparison of Padova2000 and BaSTI stellar evolution 
models by using Vazdekis/MILES based SSP libraries.

\subsection{Effects of stellar evolution isochrones}
We select Lib1 and Lib3 libraries to check the relative parameter offsets 
($\Delta P = P_{\rm Lib3}-P_{\rm Lib1}$) caused by Padova2000 and BaSTI isochrones,
with the same setup of empirical MILES library and Salpeter IMF. 

\begin{figure*}
\centering
\includegraphics[angle=0.0,scale=0.8,origin=lb]{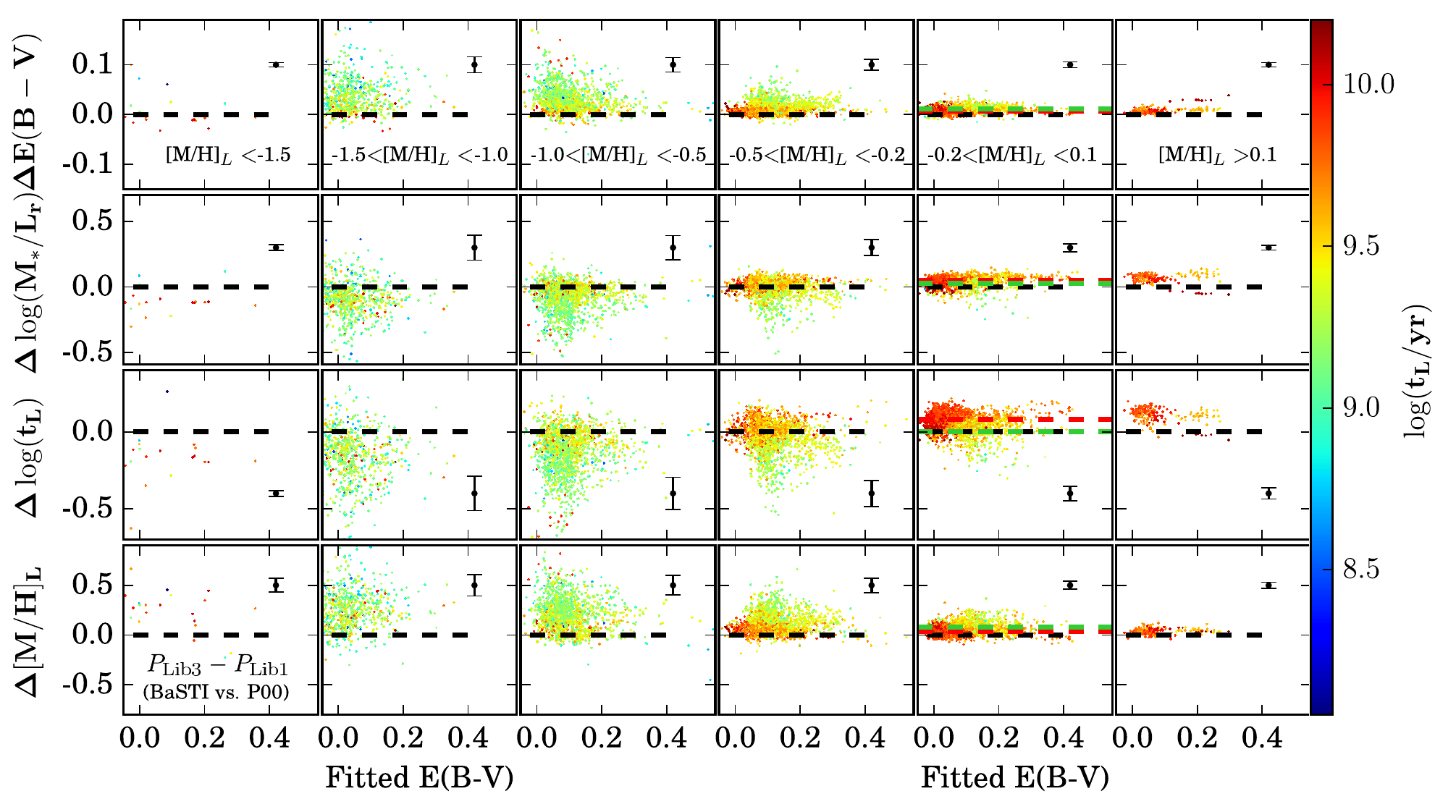}
\caption{
The relative parameter offsets ($P_{\rm Lib3}-P_{\rm Lib1}$) between the Padova2000 and BaSTI isochrones
in the case of MaNGA spectral fitting at S/N=60.
In this comparison, Lib1 and Lib3 have the same Vazdekis/MILES model and Salpeter IMF.
Lines, points, error bars, and colors are the same as described in Figure \ref{miles_bc03_fit}.
In the fifth column, which contains the largest number of data points, we plot the red and green 
dashed lines to show the median offsets ($\Delta$E(B-V), $\Delta \log(M_*/L_r)$, $\Delta \log t_L$, 
and $\Delta$[M/H]$_L$) of those old ($t_L>3$ Gyr) and young ($t_L<3$ Gyr) spectra, respectively.
}
\label{p00_basti_obs}
\end{figure*}

In Figure \ref{p00_basti_obs} we plot the relative parameter offsets derived from $P_{\rm Lib3}-P_{\rm Lib1}$.
These offsets can be summarized as follows:
\begin{itemize}
	\item Compared to the fitting with pPXF+Lib1, fittings with pPXF+Lib3 have a maximum $\sim 0.1$ magnitude larger E(B-V) offsets for those younger spectra ($t_L<10^{9.5}$yr) in the metallicity bins $-1.5<$[M/H]$_L<-0.2$. 
Spectra with older ages ($t_L>10^{9.5}$yr) tend to have smaller E(B-V) offset.
	\item The measured relative light-weighted ages and $M_*/L_r$ offsets tend to be positive for those old spectra, but negative for those young spectra.
	\item The fitted relative [M/H]$_L$ offsets increase with younger ages.
\end{itemize}

For the stellar evolution model analyses, it is hard to predict
which one is closer to the real stellar evolution. However, we can check which stellar evolution model
is closer to the reality by designing a contrast test. For example, if we assume the Padova2000 
model is closer to the reality, then if we generate mock spectra using Padova2000 and then fit it with 
both Lib1 and Lib3, it should yield similar relative parameter offset ($\Delta P = P_{\rm Lib3}-P_{\rm Lib1}$) 
as that yielded by the real spectra as shown in Figure \ref{p00_basti_obs}. Otherwise, if the observed 
relative parameter offset trends are similar to that based on the mock spectra generated using Lib3, 
then we can conclude that the BaSTI model describes the stellar evolution of local galaxies better.

\begin{figure*}
\centering
\includegraphics[angle=0.0,scale=0.8,origin=lb]{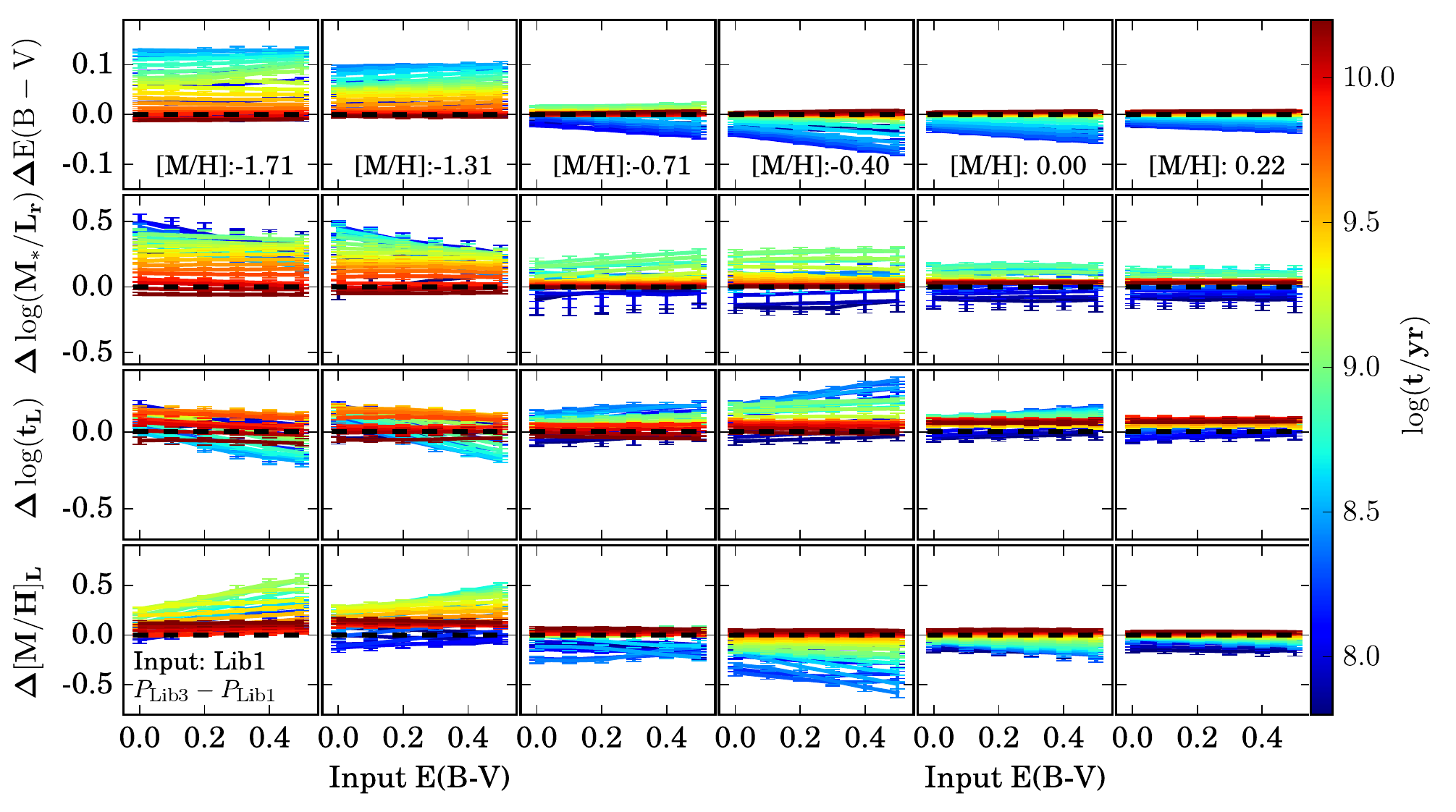}\\
\includegraphics[angle=0.0,scale=0.8,origin=lb]{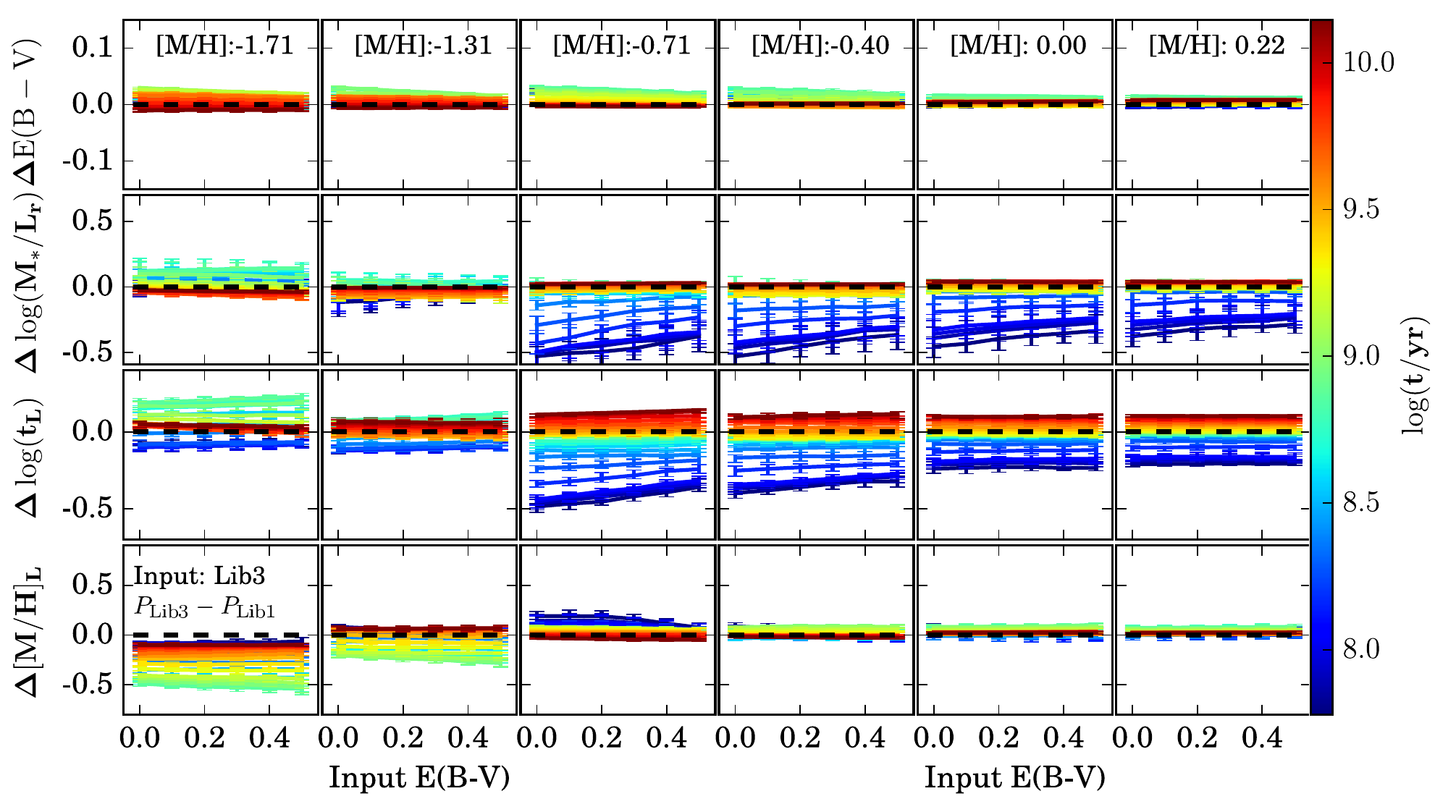}
\caption{
$Top:$ The mock spectra are generated based on the Lib1 library, then spectral fittings are 
performed with pPXF+Lib1 and pPXF+Lib3. The relative parameter offsets ($P_{\rm Lib3}-P_{\rm Lib1}$) 
in the simulation can be compared with that shown in Figure \ref{p00_basti_obs}.
$Bottom:$ The mock spectra are generated based on Lib3 library, then the spectral fitting and 
analyses of relative parameter offsets are performed in the same way as done in the $Top$ figure.
Lines, points, and colors are the same as described in Figure \ref{miles_imf}.
}
\label{p00_basti_simu}
\end{figure*}
\begin{figure*}
\centering
\includegraphics[angle=0.0,scale=0.9,origin=lb]{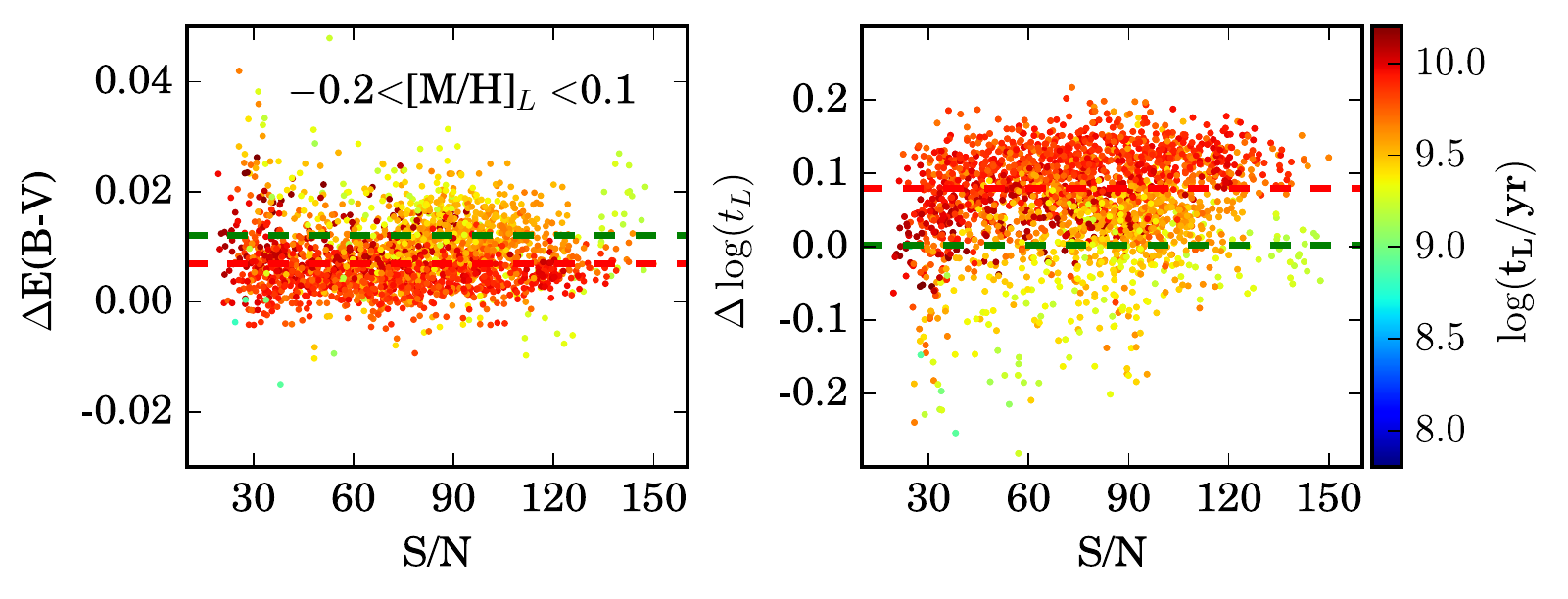}
\caption{Variations of $\Delta$E(B-V) (left panel) and $\Delta \log(t_L)$ (right panel) with S/N 
for the $-0.2<$[M/H]$_L<0.1$ bin. The points are colored by the luminosity-weighted
stellar ages as shown in the right colorbar. The red and green lines shown in each panel are defined the
same as shown in Figure \ref{p00_basti_obs}.
}
\label{snr_isochrone}
\end{figure*}

Firstly, we assume that the Padova2000 model is closer to the reality than the BaSTI model, and generate the 
mock spectra based on Lib1 SSP library. Whether this assumption is correct can be confirmed by comparing the 
trends of parameter offsets derived from both simulation (Figure \ref{p00_basti_simu} $top$) and observation 
(Figure \ref{p00_basti_obs}).
After the comparison we can see that only the $\Delta$E(B-V) and $\Delta$[M/H]$_L$ at low metallicity cases ([M/H]$_L<-1.0$) have
similar trends for the simulation and observation tests. When [M/H]$_L>-1.0$, the parameter offset trends
of the four parameters in the simulation are opposite to the observation,
which means that the current assumption is not reasonable for richer metallicities. 

Secondly, we assume the BaSTI model is closer to the reality, the mock spectra are then generated 
using Lib3. In the bottom panels of Figure \ref{p00_basti_simu},
the E(B-V) offset trends are the same as shown in Figure \ref{p00_basti_obs}. The only difference
is that the results are more scattered in the observational results. For those spectra with 
[M/H]$_L>-1.0$, the $\Delta \log(M_*/L_r)$, $\Delta \log(t_L/\rm yr)$, and $\Delta$[M/H]$_L$ 
show similar offset trends at different ages. For example, in the fifth metallicity bin ($-0.2<{\rm [M/H]}_L<0.1$) 
of Figure \ref{p00_basti_obs}, the red ($t>3$ Gyr) and green ($t<3$ Gyr) dashed lines show similar offset 
trends as shown in Figure \ref{p00_basti_simu} $bottom$. However, in low metallicity bins ([M/H]$_L<-1.0$),
the trends from simulation and observation tests are opposite.

We also explore the relations between parameter offsets and spectral S/N's at the $-0.2<$[M/H]$_L<0.1$ 
bin in Figure \ref{snr_isochrone}. From both $\Delta$E(B-V) vs. S/N in the left panel and $\Delta \log(t_L)$ vs. S/N
in the right panel, we can see that at smaller S/N ($<60$), the parameter offsets are actually contaminated 
by the S/N. With increasing S/N, the $\Delta \log(t_L)$ differences between the old and young spectra 
become more clear. At S/N$\ge 60$, the offset trends are more likely that shown in the bottom panels of 
Figure \ref{p00_basti_simu}.

Based on the above analyses, we conclude that the Padova2000 model matches the local galaxies better than the BaSTI model at [M/H]$_L<-1.0$,
while the BaSTI model is closer to the local galaxy evolution than Padova2000 model at [M/H]$_L>-1.0$.

It would be helpful for improving the modeling of stellar evolution isochrones by finding out the detailed differences 
between the Padova2000 and BaSTI models, and searching for the dominant reason of the above conclusion. However, in this paper,
we only present the comparison results and leave the detailed analyses of the two models in future works.

\section{Discussion}
\subsection{Basic assumptions}
There are two basic assumptions in this paper: 1) a universal form of log-normal SFH, and 2) the relative
parameter offset is calculated by assuming that one of the selected five SSP libraries is closer to real observations.

From Figures 6, 8, 9, and 10 we find that parameter offsets derived from
real spectra are more scattered than the mock ones. This might be caused by two reasons: 
(1) our assumed log-normal SFH in the simulated spectra does not accurately describe the real SFHs in local galaxies, 
(2) none of the five selected SSP libraries can represent the reality exactly. 

As to the relative parameter offsets, our comparison is based on the fact that there are only small 
parameter offsets if an SSP library is close to the real case. If there are larger difference for 
both the two SSP libraries, we cannot derive reasonable values of those relative parameter offsets 
based on the simulation. 

\subsection{Comparison with previous works}
In \cite{GC2010}, they fitted spectra of star clusters and found that the BC03 library can 
under-estimate the metallicity by 0.6 dex. In our work, we find a $\Delta M_*/L_r\sim -0.15$ dex 
offset when we compare the BC03 model to the Vazdekis/MILES model in MaNGA galaxies. 
Our $0.15$ dex relative offset is probably due to the lack of metal-rich stars
in the STELIB library compared to the MILES library. The offset obtained in \cite{GC2010} is the 
absolute bias, which may be affected by other reasons, such as the model bias and selection of isochrones.
Another thing should be noticed here is that we use the spatially resolved spectra in MaNGA survey,
which consist of composite stellar populations. This composition lead to a decrease of 
the light-weighted [M/H] offsets.

In some previous works \citep[e.g.][]{Bernardi2010}, the $M_*/L$ offset between the Chabrier IMF and Salpeter IMFs is set as 0.25 dex
to convert the stellar mass derived from the two IMFs. One thing should be noticed is that the $M_*/L$ 
offset varies with stellar age. Those young spectra can have the $M_*/L_r$ offsets smaller by up to $\sim 0.03$ dex than those old spectra. 
When we derive the $M_*/L$ from full-spectrum fitting, the $M_*/L$ offsets caused by different IMFs can also be smaller
than the intrinsic ones due to the SSP template difference and age-metallicity degeneracy effects as described in Section \ref{sec_imfvar}.

\section{Conclusions}
In this paper, we study model uncertainties of the three ingredients used for generating an SSP library: IMF, empirical stellar spectral
library, and stellar evolution model/isochrones. For each ingredient, we focus on two popular options, 
including the Salpeter and Chabrier IMFs, the empirical STELIB and MILES libraries, and the Padova2000 and BaSTI stellar
evolutionary isochrones. 

In total, we select five SSP libraries, and take two libraries for comparison at each time. 
By applying the pPXF code and different SSP libraries to MaNGA galaxies, we can derive the 
relative parameter offsets caused by different libraries. When focusing on only one ingredient variation
while keeping the other two ingredients the same, we can easily find out the parameter offsets 
caused by the selected ingredient uncertainty. To further understand the origins of these offsets, 
we perform simulations and interpret the parameter offsets based on those mock spectra with well known 
input parameters.

By using 100 randomly selected MaNGA galaxies observed with 127-fibers bundle, 
we derive the parameter offsets caused by the variation of each ingredient. By generating mock spectra 
based on assumed log-normal SFH and MEH,  we perform the same fitting and investigate the dominate 
reason of the offsets.

The SSP libraries based on the Vazdekis/MILES model introduce significant improvements on the reduced 
$\chi^2$ distributions than those based on the Galaxev/STELIB model, which might means a better 
fundamental space coverage can match the observed spectra better. From our comparison, the $\chi^2$ 
improvement mainly happens to the metal rich case with [M/H]$_L>-0.2$. Based on our detailed analyses, 
the parameter offsets at the over-solar metallicity bin are caused by differences of the relative 
chemical abundances or stellar population synthesis code improvements, instead of the age-metallicity
degeneracy or S/N effects. In the [M/H]$_L>0.1$ bin, the Galaxev/STELIB based model can introduce 
$\sim 0.15$ dex lower [M/H]$_L$, $\sim 0.19$ dex older stellar ages, and $\sim 0.12$ dex larger 
$\log(M_*/L_r)$ than those with Vazdekis/MILES based libraries.

The $M_*/L_r$ derived from the Chabrier IMF are of course systematically lower than that from 
the Salpeter IMF by a median of $\sim 0.22$ dex from the Galaxev/STELIB model, but $\sim 0.15$ dex 
from the Vazdekis/MILES model. When one fit a spectrum with different IMFs, the parameter offsets 
not only appear to $M_*/L$, but also age and metallicity. If we apply the Vazdekis/MILES based 
libraries for spectral fitting, the offsets in fitted stellar ages and metallicities 
increase with the age of the populations. However, when applying the Galaxev/STELIB model 
based libraries for test, due to their less complete fundamental parameter coverage
than the MILES library, the relative parameter offsets have no metallicity dependence.

With the MaNGA spectra, the variation of stellar evolutionary isochrones introduces no improvements 
in the reduced $\chi^2$ distributions. 
However, significant differences appear when comparing parameters derived with the two isochrones.
To check the origin of parameter offsets caused by the Padova2000 and BaSTI isochrones, we generate 
mock spectra based on both Lib1 and Lib3 to analyze the relative parameter offsets, and identify 
which one is similar to the result on observed spectra. We find that the Padova2000 isochones works 
better at low metallicity cases ([M/H]$_L<-1.0$) than the BaSTI one, while the BaSTI model matches 
the local galaxies better at higher metallicity cases ([M/H]$_L>-1.0$) than the Padova2000 one. 
With increasing spectral S/N's, the evidence becomes more robust.

For the spectral fitting of local galaxies, we would suggest to select the empirical MILES library 
to avoid the measurement offsets against rich metallicities, and the BaSTI isochrone due to its 
closer match to the local galaxy evolution. If a galaxy has potential radial IMF variation, 
spectral fitting with a fixed IMF as a function of radius will introduce age-dependent offsets of
$M_*/L$, age, and metallicity when the Vazdekis/MILES model based SSP libraries are applied.

Finally, we emphasize that it will be important to expand high-resolution stellar libraries, 
such as the ongoing MaNGA Stellar Library \citep[MaStar,][]{Yan2018}.

\section*{Acknowledgements}
We would like to thank to the anonymous referee for the suggestions that helped to improve this paper.
This work is supported by the National Key Basic Research and Development Program of China (No. 2018YFA0404501 to SM), 
by the National Natural Science Foundation of
China (NSFC) under grant number 11473032 (JG), 11333003 and 11761131004 (SM), 11390372 (SM, YL),
and 11690024 (YL), and by the National Key Program for Science
and Technology Research and Development (Grant No. 2016YFA0400704 to YL).
MC acknowledges support from a Royal Society University Research Fellowship.
RY acknowledges support by National Science Foundation grant AST-1715898.

This work makes use of data from SDSS-IV. Funding for SDSS has been provided by the Alfred P.~Sloan Foundation and 
Participating Institutions. Additional funding towards SDSS-IV has been provided by the U.S. Department of Energy 
Office of Science. SDSS-IV acknowledges support and resources from the Centre for High-Performance Computing at the 
University of Utah. The SDSS web site is \url{www.sdss.org}.

SDSS-IV is managed by the Astrophysical Research Consortium for the Participating Institutions of the 
SDSS Collaboration including the  Brazilian Participation Group, the Carnegie Institution for Science, 
Carnegie Mellon University, the Chilean Participation Group, the French Participation Group, Harvard-Smithsonian 
Center for Astrophysics, Instituto de Astrof\'isica de Canarias, The Johns Hopkins University, Kavli Institute 
for the Physics and Mathematics of the Universe (IPMU) / University of Tokyo, Lawrence Berkeley National Laboratory, 
Leibniz Institut f\"ur Astrophysik Potsdam (AIP),  Max-Planck-Institut f\"ur Astronomie (MPIA Heidelberg), 
Max-Planck-Institut f\"ur Astrophysik (MPA Garching), Max-Planck-Institut f\"ur Extraterrestrische Physik (MPE), 
National Astronomical Observatory of China, New Mexico State University, New York University, University of Notre Dame, 
Observat\'ario Nacional / MCTI, The Ohio State University, Pennsylvania State University, Shanghai Astronomical 
Observatory, United Kingdom Participation Group, Universidad Nacional Aut\'onoma de M\'exico, University of Arizona, 
University of Colorado Boulder, University of Oxford, University of Portsmouth, University of Utah, University of 
Virginia, University of Washington, University of Wisconsin, Vanderbilt University, and Yale University.

The MaNGA data used in this work is part of SDSS data release 13 \citep{albareti2017}, publicly available 
at {\tt http://www.sdss.org/dr13/manga/manga-data/}.




\begin{thebibliography}{99}
\bibitem[\protect\citeauthoryear{Albareti et al.}{2017}]{albareti2017}Albareti, F. D., Allende Prieto, C., Almeida, A. et al. 2017, ApJS, 233, 25
\bibitem[\protect\citeauthoryear{Auger et al.}{2010}]{Auger2010}Auger, M. W., Treu, T., Gavazzi, R., et al. 2010, ApJL, 721, L163
\bibitem[\protect\citeauthoryear{Baraffe et al.}{1998}]{Baraffe1998}Baraffe, I., Chabrier, G., Allard, F., Hauschildt, P. H. 1998. A\&A, 337, 403
\bibitem[\protect\citeauthoryear{Bastian, Covey \& Meyer}{2010}]{Bastian2010}Bastian, N., Covey, K. R. \& Meyer, M. R. 2010, ARA\&A, 48, 339
\bibitem[\protect\citeauthoryear{Bell \& de Jong}{2001}]{belldeJong2001}Bell, E.F. \& de Jong R. S. 2001, ApJ, 550, 212
\bibitem[\protect\citeauthoryear{Bernardi et al.}{2010}]{Bernardi2010}Bernardi, M., Shankar, F., Hyde, J. B. et al. 2010, MNRAS, 404, 2087
\bibitem[\protect\citeauthoryear{Bershady et al.}{2010}]{Bershady2010}Bershady, M. A., Verheijen, M. A. W., Swaters, R. A., et al. 2010, ApJ, 716, 198
\bibitem[\protect\citeauthoryear{Bertelli et al.}{1994}]{bertelli1994}Bertelli, G., Bressan, A., Chiosi, C., Fagotto, F., Nasi, E. 1994. A\&AS, 106, 275
\bibitem[\protect\citeauthoryear{Blanc et al.}{2013}]{Blanc2013}Blanc, G. A., Schruba, A., Evans, N. J., II, et al. 2013, ApJ, 764, 117
\bibitem[\protect\citeauthoryear{Bressan, Chiosi \& Fagotto}{1994}]{bcf1994}Bressan A, Chiosi C, Fagotto F. 1994. ApJS, 94, 63
\bibitem[\protect\citeauthoryear{Brodie et al.}{2014}]{Brodie2014}Brodie, J. P., Romanowsky, A. J., Strader, J., et al. 2014, ApJ, 796, 52
\bibitem[\protect\citeauthoryear{Bruzual}{1983}]{bruzual1983}Bruzual, G. 1983, ApJ, 273, 105
\bibitem[\protect\citeauthoryear{Bruzual \& Charlot}{1993}]{bc1993}Bruzual, G. \& Charlot, S. 1993, ApJ, 405, 538
\bibitem[\protect\citeauthoryear{Bruzual \& Charlot}{2003}]{bc2003}Bruzual, G. \& Charlot, S. 2003, MNRAS, 344, 1000
\bibitem[\protect\citeauthoryear{Bundy et al.}{2015}]{Bundy2015}Bundy, K., Bershady, M. A., Law, D. R. et al. 2015, ApJ, 798, 7
\bibitem[\protect\citeauthoryear{Calzetti et al.}{2000}]{Calzetti2000}Calzetti, D., Armus, L., Bohlin, R. C. et al. 2000, ApJ, 533, 682
\bibitem[\protect\citeauthoryear{Cappellari \& Copin}{2003}]{CC2003}Cappellari, M. \& Copin, Y. 2003, MNRAS, 342, 345
\bibitem[\protect\citeauthoryear{Cappellari \& Emsellem}{2004}]{CappellariEmsellem2004}Cappellari, M. \& Emsellem, E. 2004, PASP, 116, 138
\bibitem[\protect\citeauthoryear{Cappellari et al.}{2011}]{Cappellari2011}Cappellari, M., Emsellem, E., Krajnovic, D., et al. 2011, MNRAS, 413, 813
\bibitem[\protect\citeauthoryear{Cappellari et al.}{2012}]{cappellari2012}Cappellari, M., McDermid, R. M., Alatalo, K. et al. 2012, Nature, 484, 485
\bibitem[\protect\citeauthoryear{Cappellari}{2016}]{Cappellari2016}Cappellari, M., 2016, ARA\&A, 54, 597
\bibitem[\protect\citeauthoryear{Cappellari}{2017}]{Cappellari2017}Cappellari, M., 2017, MNRAS, 466, 798
\bibitem[\protect\citeauthoryear{Cassisi et al.}{2000}]{Cassisi2000}Cassisi, S., Castellani, V., Ciarcelluti, P., Piotto, G., \& Zoccali, M. 2000, MNRAS, 315, 679
\bibitem[\protect\citeauthoryear{Chabrier \& Baraffe}{1997}]{CB1997}Chabrier, G. \& Baraffe, I. 1997, A\&A, 327, 1039
\bibitem[\protect\citeauthoryear{Chabrier}{2003}]{Chabrier2003}Chabrier, G. 2003, PAPS, 115, 763
\bibitem[\protect\citeauthoryear{Charlot \& Bruzual}{1991}]{cb1991}Charlot, S. \& Bruzual, A. G. 1991, ApJ, 367, 126
\bibitem[\protect\citeauthoryear{Chen et al.}{2010}]{Chen2010}Chen, X. Y., Liang, Y. C., Hammer, F. et al. 2010, A\&A, 2010, 515, A101
\bibitem[\protect\citeauthoryear{Chen et al.}{2011}]{chen2011}Chen, Y., Trager, S., Peletier, R., Lancon, A. 2011. Journal of Physics Conference Series, 328:012023
\bibitem[\protect\citeauthoryear{Cid Fernandes et al.}{2005}]{cid2005}Cid Fernandes, R., Mateus, A., Sodre, L., Stasinska, G., Gomes, J. M. 2005, MNRAS, 358, 363
\bibitem[Cid Fernandes(2018)]{cid2018} Cid Fernandes, R.\ 2018, \mnras, 480, 4480 
\bibitem[\protect\citeauthoryear{Cioni et al.}{2006}]{Cioni2006}Cioni, M.-R. L., Girardi, L., Marigo, P., \& Habing, H. J. 2006, A\&A, 448, 77
\bibitem[\protect\citeauthoryear{Conroy}{2013}]{conroy2013}Conroy, C. 2013, ARA\&A, 51, 393
\bibitem[\protect\citeauthoryear{Conroy, Gunn, \& White}{2009}]{cgw2009}Conroy, C., Gunn, J. E. \& White, M. 2009, ApJ, 699, 486
\bibitem[\protect\citeauthoryear{Conroy \& Gunn}{2010}]{CG2010}Conroy, C. \& Gunn, J. E. 2010, ApJ, 712, 833
\bibitem[\protect\citeauthoryear{Conroy \& van Dokkum}{2012}]{cv2012}Conroy, C., \& van Dokkum, P. G. 2012, ApJ, 760, 71
\bibitem[\protect\citeauthoryear{Cordier et al.}{2007}]{cordier2007}Cordier, D., Pietrinferni, A., Cassisi, S., Salaris, M. 2007, AJ, 133, 468
\bibitem[\protect\citeauthoryear{Croom et al.}{2012}]{Croom2012}Croom, S. M., Lawrence, J. S., Bland-Hawthorn, J., et al. 2012, MNRAS, 421, 872
\bibitem[\protect\citeauthoryear{de Zeeuw et al.}{2002}]{deZeeuw2002}de Zeeuw, P. T., Bureau, M., Emsellem, E., et al. 2002, MNRAS, 329, 513
\bibitem[\protect\citeauthoryear{Dotter et al.}{2007}]{Dotter2007}Dotter, A., Chaboyer, B., Ferguson, J. W., Lee, H.-C., Worthey, G., Jevremovic, D., \& Baron, E. 2007, ApJ, 666, 403
\bibitem[\protect\citeauthoryear{Dotter et al.}{2008}]{Dotter2008}Dotter, A., Chaboyer, B., Jevremovic, D. et al. 2008. ApJS, 178, 89
\bibitem[\protect\citeauthoryear{Fioc \& Rocca-Volmerange}{1997}]{frv1997}Fioc, M. \& Rocca-Volmerange, B. 1997, A\&A, 326, 950
\bibitem[\protect\citeauthoryear{Ge et al.}{2018}]{ge2018}Ge, J. Q., Yan, R. B, Cappellari, M. et al. 2018, MNRAS, 478, 2633
\bibitem[\protect\citeauthoryear{Girardi et al.}{2000}]{girardi2000}Girardi, L., Bressan, A., Bertelli, G., Chiosi, C. 2000. A\&AS 141, 371
\bibitem[\protect\citeauthoryear{Gladders et al.}{2013}]{Gladders2013}Gladders, M. D., Oemler, A., Dressler, A. et al. 2013, ApJ, 770, 64
\bibitem[\protect\citeauthoryear{Gonzales Delgado \& Cid Fernandes}{2010}]{GC2010}Gonzales Delgado, R. \& Cid Fernandes, R. 2010, MNRAS, 2010, 403, 797
\bibitem[\protect\citeauthoryear{Gregg et al.}{2006}]{gregg2006}Gregg, M. D., Silva, D., Rayner, J., Worthey, G., Valdes, F., et al. 2006. In The 2005 HST Calibration
Workshop: Hubble After the Transition to Two-Gyro Mode, eds. Koekemoer, A. M., Goudfrooij, P., Dressel, L. L.
\bibitem[\protect\citeauthoryear{Gunn \& Stryker}{1983}]{gs1983}Gunn, J. E. \& Stryker, L. L. 1983. ApJS, 52, 121
\bibitem[\protect\citeauthoryear{Heap \& Lindler}{2011}]{hl2011}Heap, S. R. \& Lindler, D. 2011. In Astronomical Society of the Pacific Conference Series, eds. Johns-
Krull, C., Browning, M. K., West, A. A., vol. 448 of Astronomical Society of the Pacific Conference Series
\bibitem[\protect\citeauthoryear{Jones}{1999}]{jones1999}Jones, B. J. T. 1999, MNRAS, 307, 376
\bibitem[\protect\citeauthoryear{Kacharov et al.}{2018}]{Kacharov2018}Kacharov, N., Neumayer, N., Seth, A.~C., et al.\ 2018, MNRAS, 480, 1973
\bibitem[\protect\citeauthoryear{Koleva et al.}{2008}]{Koleva2008}Koleva, M., Prugniel, Ph. Ocvirk, P. et al. 2008, MNRAS, 385, 1998
\bibitem[\protect\citeauthoryear{Kroupa}{2001}]{Kroupa2001}Kroupa, P. 2001, MNRAS, 322, 231
\bibitem[\protect\citeauthoryear{La Barbera et al.}{2013}]{LaBarbera2013}La Barbera, F., Ferreras, I., Vazdekis, A., et al. 2013, MNRAS, 433, 3017
\bibitem[\protect\citeauthoryear{La Barbera et al.}{2016}]{LaBarbera2016}La Barbera F., Vazdekis A., Ferreras I. et al., 2016, MNRAS, 457, 1468
\bibitem[\protect\citeauthoryear{Lasker et al.}{2013}]{Lasker2013}Lasker, R., van den Bosch, R. C. E., van de Ven, G., et al. 2013, MNRAS, 434, L31
\bibitem[\protect\citeauthoryear{Le Borgne et al.}{2003}]{leborgne2003}Le Borgne, J. F., Bruzual, G., Pello, R., Lancon, A., Rocca-Volmerange, B., et al. 2003. A\&A,
402, 433
\bibitem[\protect\citeauthoryear{Leitherer et al.}{1999}]{leitherer1999}Leitherer, C. et al. 1999. ApJS, 123, 3
\bibitem[\protect\citeauthoryear{Lejeune, Cuisinier \& Buser}{1997}]{Lejeune1997}Lejeune, T., Cuisinier, F. \& Buser, R. 1997, A\&AS, 125, 229
\bibitem[\protect\citeauthoryear{Lejeune, Cuisinier \& Buser}{1998}]{Lejeune1998}Lejeune, T., Cuisinier, F. \& Buser, R. 1998, A\&AS, 130, 65
\bibitem[\protect\citeauthoryear{Li et al.}{2017}]{li2017}Li, H. Y., Ge, J. Q., Mao, S., Cappellari, M. et al. 2017, ApJ, 838, 77
\bibitem[\protect\citeauthoryear{Ma et al.}{2014}]{Ma2014}Ma, C.-P., Greene, J. E., McConnell, N., et al. 2014, ApJ, 795, 158
\bibitem[\protect\citeauthoryear{Maraston}{1998}]{maraston1998}Maraston, C. 1998, MNRAS, 300, 872
\bibitem[\protect\citeauthoryear{Maraston}{2005}]{Maraston2005}Maraston, C. 2005, MNRAS, 362, 799
\bibitem[\protect\citeauthoryear{Maraston \& Stromback}{2011}]{ms2011}Maraston, C. \& Stromback, G. 2011, MNRAS, 418, 2785
\bibitem[\protect\citeauthoryear{Marigo et al.}{2008}]{marigo2008}Marigo, P., Girardi, L., Bressan, A., Groenewegen, M.A.T., Silva, L., Granato, G.L. 2008. A\&A, 
482, 883
\bibitem[\protect\citeauthoryear{Martins et al.}{2005}]{Martins2005}Martins, L. P., Gonzalez Delgado, R. M. Leitherer, C. et al. 2005, MNRAS, 358, 49
\bibitem[\protect\citeauthoryear{Meynet \& Maeder}{2000}]{mm2000}Meynet, G. \& Maeder, A. 2000, A\&A, 361, 101
\bibitem[\protect\citeauthoryear{Pacifici et al.}{2016}]{Pacifici2016}Pacifici, C., Kassin, S. A., Weiner, B. J. et al. 2016, 832, 79
\bibitem[\protect\citeauthoryear{Parikh et al.}{2018}]{Parikh2018}Parikh, T., Thomas, D. Maraston, C. et al. 2018, MNRAS, 477, 3954
\bibitem[\protect\citeauthoryear{Pickles}{1998}]{pickles1998}Pickles, A. J. 1998. PASP, 110, 863
\bibitem[\protect\citeauthoryear{Pietrinferni et al.}{2004}]{pietrinferni2004}Pietrinferni, A., Cassisi, S., Salaris, M., Castelli, F. 2004, ApJ, 612, 168
\bibitem[\protect\citeauthoryear{Posacki et al.}{2015}]{Posacki2015}Posacki, S., Cappellari, M., Treu, T. et al. 2015, MNRAS, 446, 493
\bibitem[\protect\citeauthoryear{Prugniel \& Soubiran}{2001}]{ps2001}Prugniel, P. \& Soubiran, C. 2001. A\&A, 369, 1048
\bibitem[\protect\citeauthoryear{Rayner et al.}{2009}]{rayner2009}Rayner, J. T., Cushing, M. C., Vacca, W. D. 2009, ApJS, 185, 289
\bibitem[\protect\citeauthoryear{Salpeter}{1955}]{salpeter1955}Salpeter, E. E. 1955, ApJ, 121, 161
\bibitem[\protect\citeauthoryear{Sanchez et al.}{2012}]{Sanchez2012}Sanchez, S. F., Kennicutt, R. C., Gil de Paz, A., et al. 2012, A\&A, 538, A8
\bibitem[\protect\citeauthoryear{Sanchez-Blazquez et al.}{2006}]{sb2006}Sanchez-Blazquez, P., Peletier, R. F., Jimenez-Vicente, J., Cardiel, N., Cenarro, A. J., et al. 2006. MNRAS, 371, 703
\bibitem[\protect\citeauthoryear{Schaller et al.}{1992}]{schaller1992}Schaller, G., Schaerer, D., Meynet, G., Maeder, A. 1992, A\&AS, 96, 269
\bibitem[\protect\citeauthoryear{Searle, Sargent \& Bagnuolo}{1973}]{ssb1973}Searle, L., Sargent, W. L. W., \& Bagnuolo, W. G., 1973, ApJ, 179, 427
\bibitem[\protect\citeauthoryear{Shetty \& Cappellari}{2015}]{sc2015}Shetty, S., Cappellari, M. 2015, MNRAS, 454, 1332
\bibitem[\protect\citeauthoryear{Schaller et al.}{1992}]{Schaller1992}Schaller, G., Schaerer, D., Meynet, G., \& Maeder, A. 1992, A\&AS, 96, 269
\bibitem[\protect\citeauthoryear{Spiniello et al.}{2012}]{Spiniello2012}Spiniello, C., Trager, S. C., Koopmans, L. V. E., \& Chen, Y. P. 2012, ApJL, 753, L32
\bibitem[\protect\citeauthoryear{Spiniello, Trager \& Koopmans}{2015}]{Spiniello2015}Spiniello, C., Trager, S. C. \& Koopmans, L. V. E., 2015, ApJ, 803, 87
\bibitem[\protect\citeauthoryear{Tinsley}{1968}]{tinsley1968}Tinsley, B. M. 1968. ApJ, 151, 547
\bibitem[\protect\citeauthoryear{Tinsley \& Gunn}{1976}]{tg1976}Tinsley, B. M. \& Gunn, J. E., 1976, ApJ, 203, 52
\bibitem[\protect\citeauthoryear{Tortora et al.}{2013}]{Tortora2013}Tortora, C., Romanowsky, A. J., \& Napolitano, N. R. 2013, ApJ, 765, 8
\bibitem[\protect\citeauthoryear{Treu et al.}{2010}]{Treu2010}Treu, T., Auger, M. W., Koopmans, L. V. E., et al. 2010, ApJ, 709, 1195
\bibitem[\protect\citeauthoryear{Valdes et al.}{2004}]{valdes2004}Valdes, F., Gupta, R., Rose, J. A., Singh, H. P., Bell, D. J. 2004, ApJS, 152, 251
\bibitem[\protect\citeauthoryear{van Dokkum \& Conroy}{2010}]{vC2010}van Dokkum, P. G. \& Conroy, C. 2010, Nature, 468, 940
\bibitem[\protect\citeauthoryear{Vandenberg \& Bell}{1985}]{VB1985}Vandenberg, D. A. \& Bell, R. A. 1985. ApJS, 58, 561
\bibitem[\protect\citeauthoryear{Vandenberg, Bergbusch \& Dowler}{2006}]{VBD2006}VandenBerg, D. A., Bergbusch, P. A. \& Dowler, P. D. 2006, ApJS, 162, 375
\bibitem[\protect\citeauthoryear{Vaughan et al.}{2018}]{Vaughan2018}Vaughan, S. P., Davies, R. L, Zieleniewski, S., Houghton, R. C. W. 2018, arXiv:1805.12551
\bibitem[\protect\citeauthoryear{Vazdekis}{1999}]{vazdekis1999}Vazdekis, A. 1999, ApJ, 513, 224
\bibitem[\protect\citeauthoryear{Vazdekis et al.}{2010}]{vazdekis2010}Vazdekis, A. et al. 2010, MNRAS, 404, 1639
\bibitem[\protect\citeauthoryear{Walcher et al.}{2011}]{walcher2011}Walcher, J., Groves, B., Budavari, T., Dale, D., 2011, Astrophy. Space Sci, 331, 1
\bibitem[\protect\citeauthoryear{Westera et al.}{2002}]{Westera2002}Westera, P., Lejeune, T., Buser, R., Cuisinier, F., Bruzual, G. 2002, A\&A, 381, 524
\bibitem[\protect\citeauthoryear{Willett et al.}{2013}]{Willett2013}Willett, K.~W., Lintott, C.~J., Bamford, S.~P., et al.\ 2013, MNRAS, 435, 2835
\bibitem[\protect\citeauthoryear{Worthey}{1994}]{worthey1994}Worthey, G. 1994, ApJS, 95, 107
\bibitem[\protect\citeauthoryear{Yan et al.}{2018}]{Yan2018}Yan, R., Chen, Y., Lazarz, D. et al. 2018, arXiv:1812.02745
\bibitem[\protect\citeauthoryear{Yi et al.}{2001}]{Yi2001}Yi, S., Demarque, P., Kim, Y.-C., Lee, Y.-W., Ree, C. H., Lejeune, T., \& Barnes, S. 2001, ApJS, 136, 417
\bibitem[\protect\citeauthoryear{Yi, Kim \& Demarque}{2003}]{Yi2003}Yi, S. K., Kim, Y. C., Demarque, P. 2003. ApJS, 144, 259
\bibitem[\protect\citeauthoryear{Zieleniewski}{2017}]{Zieleniewski2017}Zieleniewski, S., Houghton, R. C. W., Thatte, N. et al. 2017, MNRAS, 465, 192
\end{thebibliography}




\appendix

\section{Simulations}\label{app_simu}

When applying the observed spectra for model uncertainty analyses, one potential problem is: 
even when we divide the spectral sample into detailed subsamples, it is hard to have a
thorough understanding of the origin of these offsets. 
We interpret these model uncertainties further using mock data.
To match the observed spectra, we need to construct the galaxy star formation history (SFH) and metallicity enrichment history (MEH).

\subsection{Construction of SFH and MEH}
Many efforts have been dedicated to the study of SFH of different kinds of galaxies, 
the log-normal SFH tends to be a simple model and have been supported by many observations 
\citep[e.g.][]{Gladders2013, Pacifici2016}.
We then assume a log-normal SFH and metal enrichment history (MEH) as follows:
\begin{equation}
    {\rm SFH} (M_*/{\rm yr}) = \frac{1}{\sqrt{2\pi}\tau_t} \exp\left[\frac{(t-t_0)^2}{2\tau_t^2}\right].
	\label{eq:sfh}
\end{equation}
\begin{equation}
    {\rm MEH} (Z_{\sun}) = \frac{1}{\sqrt{2\pi}\tau_Z} \exp\left[\frac{(Z-Z_0)^2}{2\tau_Z^2}\right].
	\label{eq:mfh}
\end{equation}
\begin{figure*}
\centering
\includegraphics[angle=0.0,scale=0.8,origin=lb]{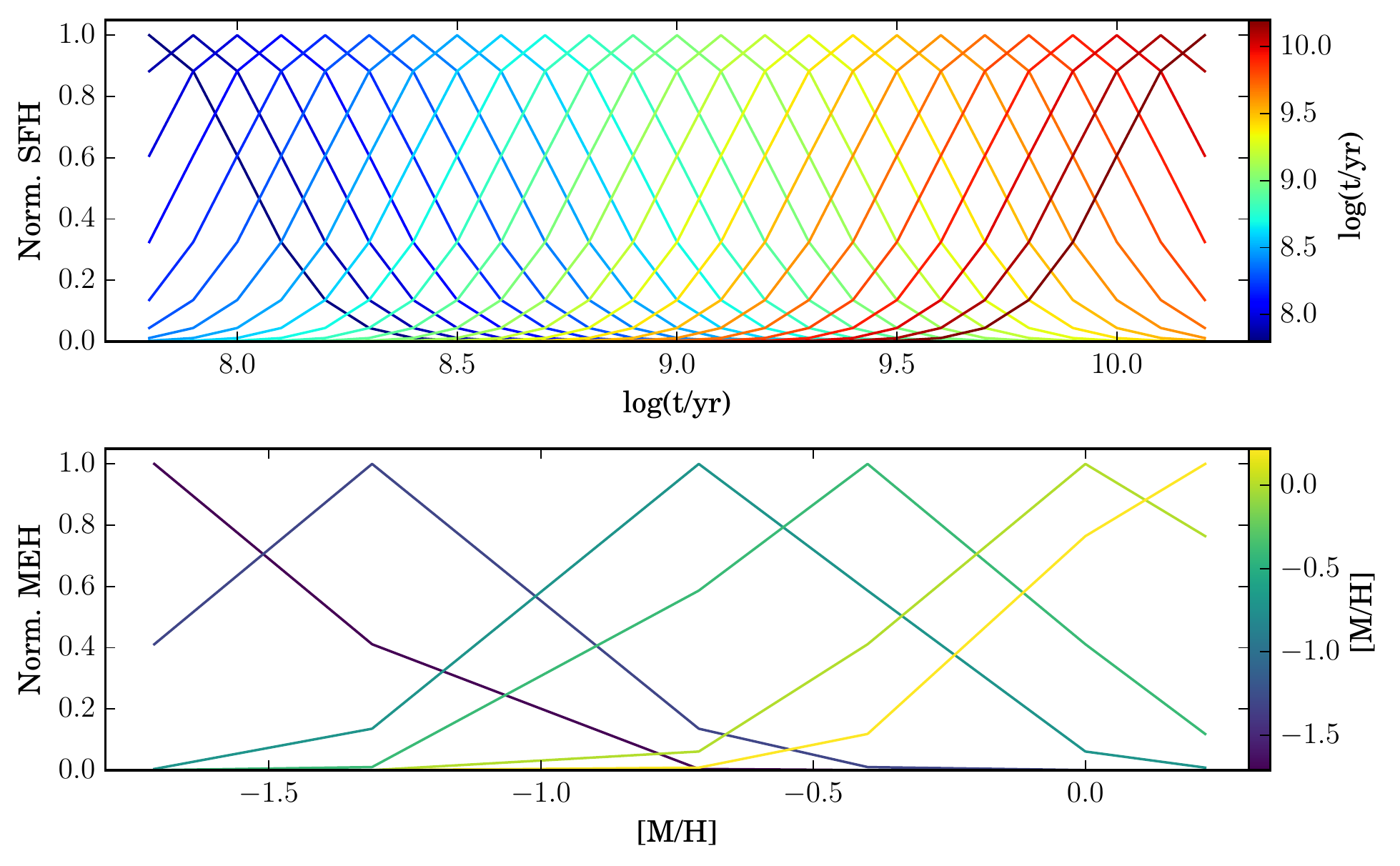}
\caption{
Log-normal star formation history (SFH, top panel) and metallicity enrichment history (MEH, bottom panel) adopted for generating mock spectra. 
The top panel shows 25 SFH curves adopted to generate mock spectra, with the peak SSP age 
shown by the right color bar (blue to red: young to old).
The bottom panel shows 6 MEHs by setting each SSP peak metallicity from poor to rich, where metallicities
are labelled with purple to yellow colors.
}
\label{sfh_mfh}
\end{figure*}

Based on our constructed SFH and MEH, Figure \ref{sfh_mfh} shows the SFH in the age space (top panel)
and metallicity space (bottom panel) by setting the color of each curve according to the peak of the log-normal distribution.

\begin{figure*}
\centering
\includegraphics[angle=0.0,scale=0.6,origin=lb]{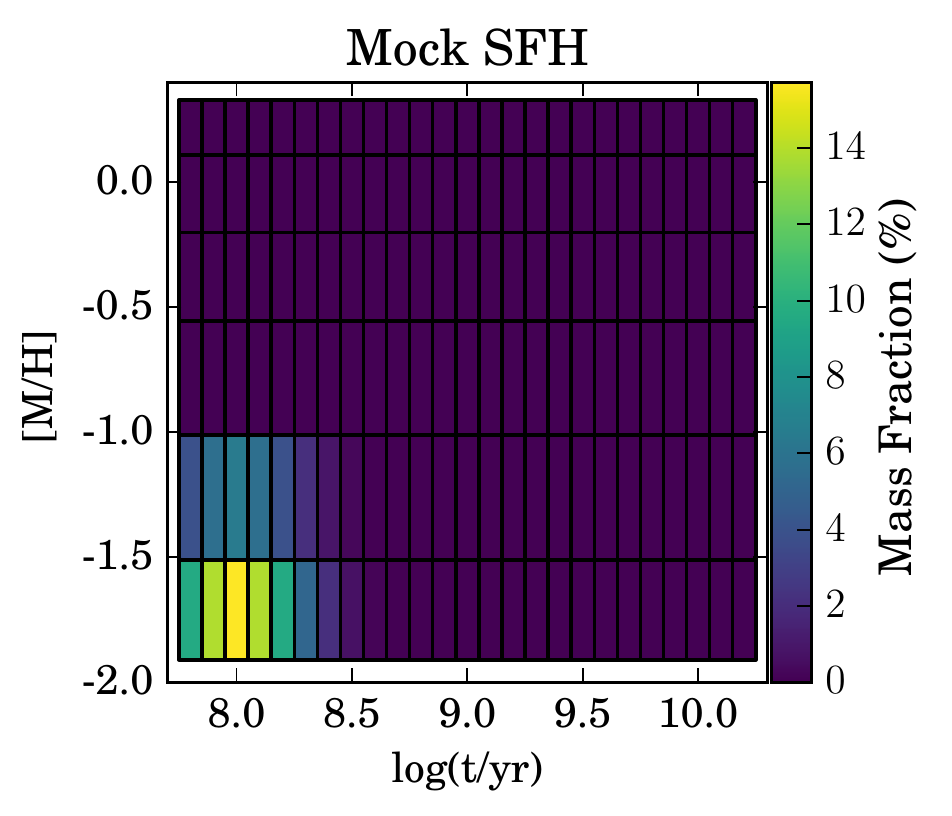}
\includegraphics[angle=0.0,scale=0.6,origin=lb]{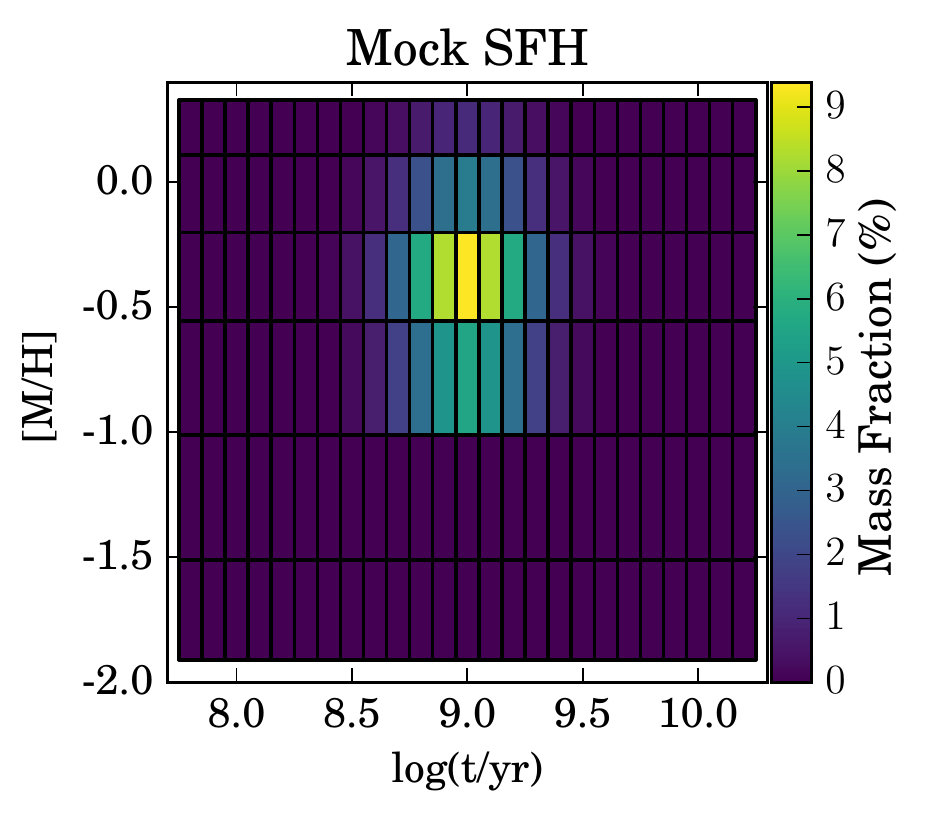}
\includegraphics[angle=0.0,scale=0.6,origin=lb]{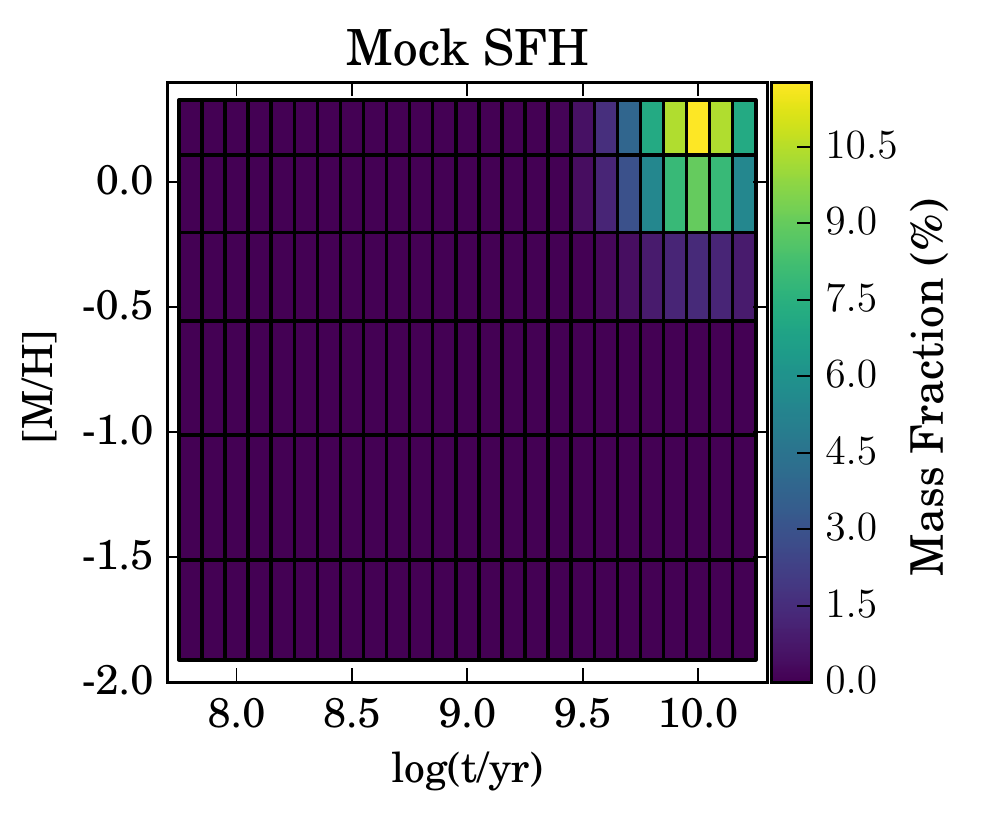}
\caption{
Three mock SFH examples with the peak age and metallicity located at [$\log (t/\rm yr)$, [M/H]]=
[8.0, $-1.71$], [9.0, $-0.4$], and [10.0, 0.22]. In each panel, the right color bar indicates the 
linear mass fraction of each SSP components.
}
\label{sfh_example}
\end{figure*}

The final SFH-MEH map in the age-metallicity parameter space can be obtained, which appears the same
as the fitted results. Here we show three examples with different age and metallicity peaks in Figure \ref{sfh_example}.

We then use the assumed SFH-MEH map to generate mock spectra, which are then used for model offset analyses.

\subsection{Mock data generation and spectral fitting}
With the constructed SFH, we first obtain the mass fraction of each SSP, then convert the corresponding mass fraction 
of each SSP to light fraction based on $M_*/L_r$. A mock spectrum is then 
generated based on the $r-$band light fraction of each SSP. According to the different wavelength 
coverages of Galaxev/STELIB and Vazdekis/MILES models, we set the spectral fitting range of mock data to 3600-7350\AA, 
which eliminates $\sim 50$\AA~ at both the blue and red end to make sure
the model SSPs have larger spectral coverage than mock spectra.

Due to the different spectral resolutions of Galaxev/STELIB (FWHM=2.76\AA) and Vazdekis/MILES (FWHM=2.54\AA), 
all the mock spectra generated based on Vazdekis/MILES libraries are smoothed to FWHM=2.76\AA. 
The MaNGA spectral resolution is similar to Galaxev/STELIB library, with a velocity scale of 69 km/s \citep{albareti2017}.
The mock spectra are then sampled in logarithmic wavelength grid, which is required by pPXF, and the velocity
scale is uniformly set to 69 km/s. To make our mock spectra more like observed ones, we artificially add a 100 km/s 
velocity dispersion to these spectra. By assuming a flat error spectrum along wavelength, we then
add noise to the velocity dispersion added mock spectra. The final generated mock spectra have S/N=60 at 
5500\AA~ (median S/N at the wavelength window[5490, 5510]\AA).

During spectral fitting, we use the full spectral information with emission-line regions masked, to 
match the fitting done to the MaNGA data.
The current version of pPXF can fit both the stellar emission and gas emission together.
In \cite{ge2018}, we avoid the emission line fitting since STARLIGHT code can only fit the stellar spectra,
here we also turn off the emission line fitting process and only focus on stellar continuum fitting.

For dust extinction curves in different galaxies, we allow the input E(B-V) to vary from 0.0 to 0.5
as done in \cite{ge2018}, by adopting the CAL \citep{Calzetti2000} dust reddening curve.

\subsection{Parameter biases of the pPXF fitting at S/N=60}\label{simu_bias}

In \cite{ge2018}, we have checked the pPXF fitting biases at S/N=60 for those single-SSP or two-components
SSPs based mock spectra. With the assumed SFH here, we do the same tests to check the parameter biases
if performing the pPXF fitting at S/N=60.

\begin{figure*}
\centering
\includegraphics[angle=0.0,scale=0.85,origin=lb]{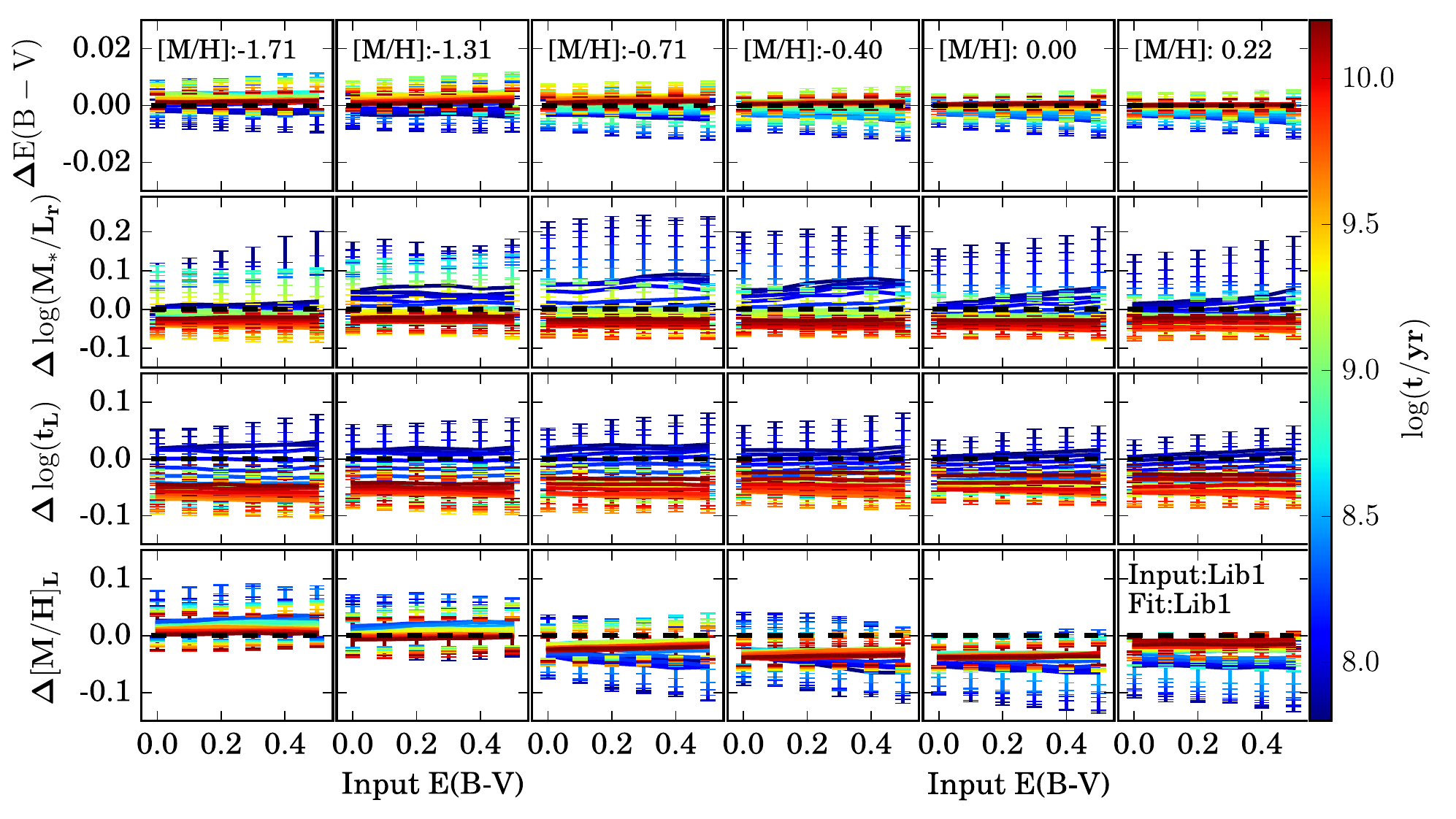}
\caption{
The fitted parameter biases of the pPXF fitting with Lib1 SSP library at S/N=60 in different metallicity bins.
We generate mock spectra based on Lib1, and then perform the fitting with pPXF+Lib1.
The relative parameter bias ($P_{\rm fit}-P_{\rm input}$) reflects the effects of age-metallicity degeneracy and
spectral S/N.
From left to right, we show the six metallicity bins ([M/H]=$-1.71$, $-1.31$, $-0.71$, $-0.4$, 0.0, 0.22) bins based 
on the peak [M/H] of each SFH. Blue to red colors represent
the stellar age ranging from 0.063 to 15 Gyr. The biases in the four parameters ($\Delta$E(B-V),
$\Delta \log(M_*/L_r)$, $\Delta \log t_L$, and $\Delta$[M/H]$_L$) are shown from top to bottom.
The zero-bias line of each parameter in each panel is labeled as the horizontal dashed line. 
The error bars indicate the 16th and 84th percentiles.
}
\label{ppxf_test}
\end{figure*}

Figure \ref{ppxf_test} shows the fitted parameter biases based on the pPXF and Lib1 SSP library.
At S/N=60, using pPXF we can constrain the dust extinction within 0.01 magnitude. 
For the recovered $\log(t_L)$ and [M/H]$_L$, all the biases and scatters in different ages have their median values 
less than 0.1 dex. The same thing happens to $M_*/L_r$, except for those spectra with stellar ages less than 0.3 Gyr,
which have larger scatters and are rare in local galaxies.

\cite{Kacharov2018} tested the pPXF fitting to spectra of the nuclear star clusters in six 
nearby galaxies. They found that even in more complex SFH cases (see their Figures 5$-$13), 
the pPXF code could still provide a good recovery of the input SFH.

%


\bsp	
\label{lastpage}
\end{document}